\documentclass[lineno]{JFM-FLM_Au}
\usepackage{graphicx}
\usepackage[dvipsnames,svgnames]{xcolor}
\usepackage{epstopdf, epsfig}
\usepackage{hyperref}
\usepackage{mathtools}
\usepackage{amsmath}
\allowdisplaybreaks[4] 
\usepackage{mathrsfs}
\usepackage{amsfonts} 
\usepackage{svg}
\usepackage{caption}
\usepackage{subcaption}
\usepackage[normalem]{ulem} 

\usepackage{titlesec}

\titleclass{\subsubsubsection}{straight}[\subsubsection]
\newcounter{subsubsubsection}[subsubsection]
\renewcommand\thesubsubsubsection{\thesubsubsection.\arabic{subsubsubsection}}

\titleformat{\subsubsubsection}
  {\normalfont\normalsize\bfseries}
  {\thesubsubsubsection}{1em}{}

\titlespacing*{\subsubsubsection}{0pt}{1.0ex}{0.6ex}

\lefttitle{Bader and Bodony}
\righttitle{Journal of Fluid Mechanics}

\title{ Beyond classical similitude: group theoretic extrapolation of hypersonic stagnation-point boundary layers}
\author{Shujaut H. Bader\aff{1} \and Daniel J. Bodony\aff{1}}
\affiliation{\aff{1}Center for Hypersonics and Entry Systems Studies, University of Illinois, Urbana-Champaign, IL, USA
}

\corresau{shbader@illinois.edu}
\if 
\usepackage{fancyhdr}
\usepackage{xcolor}
\usepackage{graphicx}

\fi 

\begin{document}
\nolinenumbers
\maketitle
\vspace{0mm}
\begin{abstract}
\textcolor{blue}{
Motivated by the need to extrapolate the results from ground-based experiments to the conditions of high-speed flight, we present the Lie equivalence symmetry analysis of the hypersonic stagnation-point boundary layers. We demonstrate the application of the equivalence symmetry on the set of coupled ODEs which are physically relevant in hypersonics. By allowing the property laws to transform along with the independent and dependent variables, the invariants derived within the similarity-reduced stagnation-point ODE formulation, identify families of non-linear maps that can be used to extrapolate the laboratory-scale predictions to flight.  In practice, implementing these maps requires the laboratory-scale ODE solution together with the lab- and flight-side thermochemical property data, which are generally available from existing databases. The maps are derived for the similarity-reduced non-dimensional temperature across the boundary layer and are shown to {collapse with independently computed  flight solutions} for a range of relevant cases.}
\end{abstract}
\begin{keywords}
\end{keywords}
\section{Introduction}
\label{sec:intro}
{\color{black}Bell X-1 broke the sound barrier in 1947 with the help of wind-tunnel tests that fell short of fully replicating the Bell X-1 flight environment due to tunnel-wall interference and a lower-than-desired Reynolds number. In 1953, the Douglas D-558-2 Skyrocket became the first aircraft to reach twice the speed of sound without fully knowing the aerodynamic loads \textit{a priori} from ground test simulation \citep{griffith_hyp_prob}. With the advent of hypersonic spaceflight which is characterized by Mach numbers, $M \geq 5$, where dissociation of air begins to become significant and heat loads become high, the ground test facilities proved to be less able to simulate the flight environment, underscoring the persistent gap between ground testing and true flight conditions.
\par 
It remains extraordinarily difficult to duplicate the full aerothermochemical environment of the stagnation region in the ground facilities at hypersonic relevant conditions. This difficulty in simultaneously reproducing flight conditions such as Mach number, Reynolds number $\mathit{Re}$, shear stress, real gas chemistry, and viscous effects in any single ground facility constitutes what can be classically termed the \textit{hypersonic simulation problem} \citep{griffith_hyp_prob}. The classical formulation posed the problem mainly in terms of globally averaged aerodynamic loads and the discrepancies between their preflight predictions and the values inferred from flight data. With rapid advances in high-speed flight in recent decades, the modern formulation of this problem targets the same underlying difficulty, now in a regime where complex aerothermochemical effects like dissociation, gas–surface interaction, radiation, and material response all play central roles in the stagnation region and boundary layer. The inclusion of additional physical effects introduces additional controlling parameters, making simultaneous matching in a single ground-test facility rarely feasible. Consequently, hypersonic research relies on a suite of complementary facilities, spanning high-enthalpy and plasma environments, to isolate specific aspects of the flight physics by matching selected nondimensional groups \citep{anabeldelval2020}. A complete characterization of real flight conditions thus requires coordinated efforts across both high-enthalpy facilities and plasma wind tunnels \citep{chazot2015high, gu2020capabilities}.
} 
\par 
In ground-based atmospheric re-entry studies, the primary objective is to reproduce the near-wall environment, namely the shear stresses, the reactive species within the boundary layer, and the resulting wall heat flux. Many such studies focus on the similitude of the key parameters to enable extrapolation from lab to flight values, i.e., matching as many non-dimensional parameters as possible. One of the earliest ground-to-flight similarity concepts is binary scaling which provides a similarity rule for the chemically nonequilibrium inviscid flow over blunt bodies using the Lighthill-Freeman ideal dissociating gas \citep{BINS1freeman1958non}. Analysis shows that in the regime where dissociation dominates recombination the equations collapse to a single similarity parameter proportional to the product of edge density and characteristic length, $\rho_e L$. Thus, two flows with the same binary scaling parameter $\rho_\text{e} L$ have similar nonequilibrium shock-layer and dissociation fields even if their individual density and length differ \citep{BINS2gibson1964similitude,BIN4muylaert1996extrapolation,BINS3inger2003generalized}. A second popular method is the concept of Local Heat Transfer Simulation (LHTS), proposed by \citet{kolesnikov1_1993conditions, kolesnikov2_2000concept}, to reproduce the highest heat load at the stagnation point of the vehicle. LHTS shows that the stagnation-point heat flux in flight can be reproduced on a test sample in a subsonic facility by matching the centerline free-stream enthalpy, stagnation pressure, and velocity gradient between the laboratory case and the hypersonic flight case. In addition, LHTS assumes that Local Thermal Equilibrium (LTE) and Local Chemical Equilibrium (LCE) hold at the boundary-layer edge of the test sample. 

Traditional extrapolation-to-flight methodologies, as outlined above, are usually formulated for experiments, with simple empirical or semi-analytical justifications; computational fluid dynamics (CFD) and data-assisted extrapolation efforts are presented in \citet{COMP2vonkaenel2009ixv}, \citet{COMP3dreyer2017multi} and \citet{COMP4fedeli2024duplication}. 
Very few mathematical or analytical efforts have been pursued beyond basic similarity theory to address the extrapolation-to-flight challenge. In this work, we leverage Lie equivalence symmetries to derive new relations that seek to link laboratory-scale observations to their flight-scale counterparts. The present work is inspired by the study of \citet{tranchant2022NEUTRON,tranchant2025generalizing}, in which Lie equivalence transformations are used to analyze {extreme} astrophysical phenomena in laboratory settings. As opposed to the standard Lie symmetries, Lie equivalence transformations scale not only the independent and dependent variables in the governing equations, but also the underlying property laws. These transformations act on the physical material elements of a model (such as constitutive or transport laws) in addition to the field variables, thereby mapping entire classes of problems into one another. The versatility of these symmetries allows them to be generalized to different systems. The method of Lie symmetry transformations has been used in many branches of physical sciences to develop particular solutions of differential equations \citep{ibragimov1994crcII}. In fluid dynamics, several studies \citep{FDLIE8oberlack1999symmetries,FDLIE1oberlack2001unified,FDLIE2lindgren2004evaluation,FDLIE3oberlack2006group, FDLIE6she2011quantifying,FDLIE7rosteck2014scaling} on plane-parallel turbulent shear flows and zero-pressure-gradient turbulent boundary layers use Lie point symmetries to derive families of invariant scaling laws for mean velocity and higher-order moments. 

\par 

All of the aforementioned fluid dynamic studies used classical Lie symmetry methods to derive scaling laws. In this work, we go one step further and, for the first time, demonstrate the applicability of Lie \textit{equivalence} symmetries in stagnation-point boundary-layer theory to obtain transformations that enable us to go beyond classical similitude. Rather than enforcing classical similitude by matching selected nondimensional groups, we adopt an equivalence-symmetry viewpoint that identifies invariant symmetry groups of the family of governing equations with \textit{arbitrary}\footnote{\textit{arbitrary} in the context of equivalence symmetry analysis means that the property laws are not \textit{fixed} in the governing equations, and that they are arbitrary functions of one of the field variables.} property laws. \textcolor{blue}{Classical similarity parameters may then emerge as special cases of invariants associated with admitted scaling transformations.
This viewpoint is consistent with the standard connection between dimensional
analysis and Lie scaling symmetries: for boundary-value problems such as the
Prandtl--Blasius boundary layer, invariance under a one-parameter Lie group of
scalings can produce the similarity reduction directly, and in this sense
scaling symmetry provides a generalization of dimensional analysis for reducing
the number of independent variables \citep{cantwell2002introduction, FDLIE6she2011quantifying}. A rich discussion on this can be found in \citet{bluman2002symmetry}.}

\textcolor{blue}{For the ease of implementation and to provide a clear proof of concept, we apply the Lie equivalence framework to the similarity-reduced stagnation-point ODEs discussed in \citet{bottin1999aerothermodynamic,barbante2001accurate} and \citet{lanza2025plasflowsolver}. We recognize that the resulting equivalence symmetry group is more restrictive than the one that would be obtained for the full stagnation-point PDE system. The ODEs are a similarity-reduced form of the same underlying laws, for which the determining equations reduce to a manageable problem while still yielding a meaningful subgroup of the full PDE equivalence and practically usable laboratory--flight mappings.}
\par
\vspace{0mm}
\section{Lie equivalence symmetries of stagnation-point ODEs}\label{sec:equiv_all_eqns}
{\color{DarkGreen}
In this section, the theoretical results from Lie equivalence symmetry analysis of the system of stagnation-point ODEs will be presented. For the paper to be self-contained, we have included the detailed derivations of the symmetries and final invariants relevant for mapping in the text below. At this point, we strongly encourage the reader to refer to the Appendix \ref{APPEND-A-LIETHEORY} for a brief introduction to the Lie-symmetry/equivalence concepts used in this study. The stagnation-point ODEs as described in \citet{lanza2025plasflowsolver} read,
\begin{subequations}\label{eq:lanzaODEs}
\begin{align}
\frac{dV}{d\eta}  &= -F, \label{eq:lanzaCont}\\
V\frac{dF}{d\eta} &= \frac{1}{2}\left(R_\rho - F^{2}\right)
+ \frac{d}{d\eta}\left(\ell_0\frac{dF}{d\eta}\right),\label{eq:lanzaMom}\\
V\frac{dg}{d\eta} &= \frac{1}{C_p}\frac{d}{d\eta}\left(\chi\frac{dg}{d\eta}\right).\label{eq:lanzaEnergy}
\end{align}
\end{subequations}
where,
\begin{align*}
    F = \frac{u}{u_e}, \quad g = \frac{T}{T_e}, \quad V = \frac{2\xi}{\partial \xi / \partial x}\left(F \frac{\partial \eta}{\partial x} + \frac{\rho v r}{\sqrt{2\xi}}\right), \quad R_\rho = \frac{\rho_\text{e}}{\rho},
\end{align*}
\begin{align*}
   \quad \xi(x) = \int_{0}^{x} \rho_e \mu_e u_e r^{2} \, dx', \quad \eta(x,y) = \frac{r u_e}{\sqrt{2\xi}} \int_{0}^{y} \rho \, dy',
\end{align*}
\begin{align*}
    l_0 = \frac{\rho \mu}{\rho_e \mu_e}, \quad \chi = \frac{\lambda \rho}{\rho_e \mu_e}
\end{align*}

Rewriting \eqref{eq:lanzaODEs} in the standard form yields,
\begin{subequations}\label{eq:RElanzaODEs}
\begin{align}
\Delta_C &\coloneqq \frac{dV}{d\eta} +F =0, \label{eq:RElanzaCont}\\
\Delta_M &\coloneqq \frac{d^2F}{d\eta^2} - \frac{1}{\ell_0} \left[ V\frac{dF}{d\eta} - \frac{1}{2} (R_\rho - F^{2}) -\frac{dF}{d\eta} \frac{d\ell_0}{dg} \frac{dg}{d\eta} \right]=0,\label{eq:RElanzaMom}\\
\Delta_E &\coloneqq \frac{d^2g}{d\eta^2} - \frac{1}{\chi} \left[ C_p V\frac{dg}{d\eta} - \left ( \frac{dg}{d\eta} \right)^2 \frac{d\chi}{dg} \right]=0.\label{eq:RElanzaEnergy}
\end{align}
\end{subequations}
The subscript `$e$' denotes the value of the variable computed at the edge of the boundary layer. For more details about the transformations used to derive the above ODEs from their underlying mass, momentum and energy PDEs, the interested reader is referred to \citet{bottin1999aerothermodynamic,barbante2001accurate} and \citet{lanza2025plasflowsolver}. It should be emphasized that the standard form of ODEs \eqref{eq:RElanzaODEs} is purely an algebraic rearrangement, introduced only to cast the system into a form that is convenient for Lie–equivalence symmetry analysis.
In the equivalence analysis of the stagnation-point ODE system, the set of elements $a_i = \{\ell_0, \chi, R_\rho,C_p\}$ depend on the non-dimensionalized temperature $g=T/T_e$ and $T_e$. For example, $\ell_0(g;T_e) = {\rho \mu}/{(\rho_e \mu_e)} = {\rho(gT_e)\mu(gT_e)}/{(\rho(T_e)\mu(T_e))}, \ \chi(g;T_e) = {\lambda \rho}/{(\rho_e \mu_e}) = {\lambda(gT_e)\rho(gT_e)}/{(\rho(T_e)\mu(T_e))}.$ Since $T_e$ does not appear explicitly in the ODEs, equivalence symmetries cannot encode its variation directly. Therefore, $T_e$ does not add any new degrees of freedom to the symmetry, and only acts as a label for the property tables from which the elements in $a_i$ are built, such that $a_i = a_i(g;T_e)$. 
Unless explicitly stated, $T_e$ in the argument of the property laws will be suppressed for brevity and these constitutive functions will be declared as $a_i(g)$.
Based on the explicit $g$-dependence of the auxiliary elements $a_i$, the system of auxiliary equations written in the standard form is,
    \begin{align}\label{eq:auxSys}
        \Delta_{a_{i,k}} := \frac{\partial a_i}{\partial x_k} = 0
    \end{align}
where $a_i \in \{R_{\rho},C_p,\ell_0,\chi\}$ and $x_k \in \{\eta, F, V\}$. These equations constrain the auxiliary functions $a_i$ with respect to $\eta,F, \text{ and } V$ while leaving the $g$-dependence unconstrained.

 {The universal infinitesimal generator and its prolongations for the equivalence transformations} of the systems of equations \eqref{eq:RElanzaODEs} and \eqref{eq:auxSys} can be written as, 
\begin{subequations}
    \begin{align}
    X &= \phi_{\eta} \partial_{\eta} + \sum^{\{F,V,g \}}_{\mathclap{d_i \in }} \phi_{d_i} \partial_{d_i} + \sum_{\mathclap{a_i \in}}^{\{R_\rho, C_p, \ell_0, \chi \}} \phi_{a_i} \partial_{a_i}, \label{eq:X}\\
    X^{(1)} &= X +  \sum_{\mathclap{d_i \in}}^{\{F,V,g \}} \phi^{(1)}_{d_i} \partial_{{d_i}^{(1)}} +   \sum_{{a_i \in}}^{\{R_\rho,C_p,\ell_0,\chi \}} \sum_{{x_j \in}}^{\{\eta,F,V,g \}} \phi^{(1)x_j}_{a_i} \partial_{{a_i}^{(x_j)}}, \label{eq:X1}\\
    X^{(2)} &= X^{(1)} +  \sum_{{d_k \in}}^{\{F,g \}} \phi^{(2)}_{d_k} \partial_{{d_k}^{(2)}}. \label{eq:X2}
\end{align}
\end{subequations}

The superscripts $(1)$ and $(2)$ on the variables $d_{i}$ and $d_k$ in $X^{(1)}$ and $X^{(2)}$ denote their first and second derivatives with respect to $\eta$, respectively. The superscript $(x_j)$ in equation for $X^{(1)}$ denotes the first derivative of the variables collected in $a_i$ with respect to the variable $x_j$. The superscripts $(1)$ and $(2)$ on the $\phi$'s in $X^{(1)}$ and $X^{(2)}$ are merely indices and do not denote derivatives of $\phi$. Any other index without parentheses is a label. When projecting the universal generator onto a given equation, only a subset of its terms is retained; the remaining terms vanish or are inapplicable for that equation and/or the auxiliary manifold.
We restrict to the equivalence group in which the infinitesimals associated with the base variables $(\eta,F,V,g)$ depend only on $(\eta,F,V,g)$, whereas the infinitesimals for the property laws $a_i\in\{\ell_0,\chi,R_\rho,C_p\}$ may depend on $(\eta,F,V,g,a_i)$; which is the standard convention in equivalence-group treatments (e.g.\ \citet{tranchant2025generalizing}).
\par
In the subsequent sections, to obtain the equivalence symmetries, we project the prolonged generator $X^{(2)}$ first onto the auxiliary equations \eqref{eq:auxSys}, and subsequently to each equation in the main system \eqref{eq:RElanzaODEs}. 

\vspace{0mm}
\subsection{Auxiliary system of equations}
Because the auxiliary system of equations contains only first derivatives of the constitutive elements, projecting the second prolongation $X^{(2)}$ onto the auxiliary manifold retains only the first-order prolongation terms  $\phi_{a_i}^{(1)x_j}$ for $x_j \in \{\eta,F,V \}$.The linearized symmetry condition (LSC) for this system reads,
 $X^{(2)} \left[\Delta_{A_{i,k}}\right] = 0,$ yielding, 
\begin{align}\label{eq:phi_aux_equals_zero}
    \phi_{a_i}^{(1)x_j} = 0, \ \forall \quad a_i \in \{R_\rho,C_p,\ell_0,\chi \} \text{ and } x_j \in \{\eta, F,V\}.
\end{align}
For $a_i \in \{R_\rho,C_p,\ell_0,\chi \}$, and $x_j \in \{\eta,F,V,g\}$, we know that, 
\begin{align}
     \phi^{(1)x_j}_{a_i}
  = D_{x_j}\!\left(\phi_{a_i}\right)
    - \sum_{x_m \in }^{\{\eta,F,V,g\}} a_{i,x_m}\,D_{x_j}\!\left(\phi_{x_m}\right).
\end{align}

To illustrate the structure of this sum, a few representative terms are written out below while keeping $a_i$ general. \\ 
\begin{subequations}
    \begin{align}
 \text{For } x_k = \eta: \notag \\
     \phi^{(1)\eta}_{a_i} &= D_{\eta}\!\left(\phi_{a_i}\right) - \frac{\partial a_i}{\partial \eta} D_{\eta}(\phi_\eta) - \frac{\partial a_i}{\partial F} D_{\eta}(\phi_F) - \frac{\partial a_i}{\partial V} D_{\eta}(\phi_V) - \frac{\partial a_i}{\partial g} D_{\eta}(\phi_g) \\
     \text{For } x_k = F: \notag \\
  \phi^{(1)F}_{a_i} &= D_{F}\!\left(\phi_{a_i}\right) - \frac{\partial a_i}{\partial \eta} D_{F}(\phi_\eta) - \frac{\partial a_i}{\partial F} D_{F}(\phi_F) - \frac{\partial a_i}{\partial V} D_{F}(\phi_V) - \frac{\partial a_i}{\partial g} D_{F}(\phi_g)  \\
   \text{For } x_k = V: \notag \\
  \phi^{(1)V}_{a_i} &= D_{V}\!\left(\phi_{a_i}\right) - \frac{\partial a_i}{\partial \eta} D_{V}(\phi_\eta) - \frac{\partial a_i}{\partial F} D_{V}(\phi_F) - \frac{\partial a_i}{\partial V} D_{V}(\phi_V) - \frac{\partial a_i}{\partial g} D_{V}(\phi_g) 
\end{align}
\end{subequations}
For any $a_i$, imposing $\partial_\eta a_i = \partial_F a_i = \partial_V a_i = 0$ on the auxiliary manifold, only the last term corresponding to $x_m=g$ survives in the sum. Therefore, these expressions reduce to,
\begin{subequations}\label{eq:aux_determining_final_step}
    \begin{align}
     \phi^{(1)\eta}_{a_i} &= D_{\eta}\!\left(\phi_{a_i}\right) - \frac{\partial a_i}{\partial g} D_{\eta}(\phi_g),  \text{ for } x_k = \eta \\
  \phi^{(1)F}_{a_i} &= D_{F}\!\left(\phi_{a_i}\right)  - \frac{\partial a_i}{\partial g} D_{F}(\phi_g),  \text{ for } x_k = F  \\
  \phi^{(1)V}_{a_i} &= D_{V}\!\left(\phi_{a_i}\right)  - \frac{\partial a_i}{\partial g} D_{V}(\phi_g),  \text{ for } x_k = V 
\end{align}
\end{subequations}
Given that the infinitesimals $\phi_{a_i} \text{ and } \phi_g$ are functions of only the base variables and arbitrary properties, their total derivatives with respect to base variables do not contain the arbitrary gradient $\partial a_i/\partial g$. Therefore a relation of the form $P(\eta,F,V,g, a_i) - ({\partial a_i}/{\partial g})\,Q(\eta,F,V,g,a_i) = 0$ must hold for all admissible $a_i$ only if $P(\eta,F,V,g, a_i) = 0, Q(\eta,F,V,g, a_i) = 0.$

Applying this argument to the expressions \eqref{eq:aux_determining_final_step}  above yields,
\begin{subequations}\label{eq:AUX_D_x_eq_zero}
    \begin{align}
    D_\eta(\phi_{a_i}) &= D_F(\phi_{a_i}) = D_V(\phi_{a_i}) = 0, \\
D_\eta(\phi_g) &= D_F(\phi_g) = D_V(\phi_g)=0.
\end{align}
\end{subequations}

In the auxiliary equations, only $a_i$ enters the system as a dependent variable, the rest of the variables $x_k\in\{\eta,F,V,g\}$ are regarded as independent coordinates, so there are no jet
variables such as $\partial_\eta F$, $\partial_\eta V$, $\partial_V \eta$, $\partial_F V$ and so on. Consequently the total
derivatives become,
\begin{align}
    D_\eta = \frac{\partial}{\partial \eta} + \frac{\partial a_i}{\partial \eta} \frac{\partial}{\partial a_i}, \ \ D_V = \frac{\partial}{\partial V} + \frac{\partial a_i}{\partial V} \frac{\partial}{\partial a_i}, \ \ D_F = \frac{\partial}{\partial F} + \frac{\partial a_i}{\partial F} \frac{\partial}{\partial a_i}.
\end{align}

Imposing the auxiliary constraints $\partial_\eta a_{i}= \partial_F a_{i}= \partial_V a_{i}=0$ reduces these total derivatives, when acting on the infinitesimals $\phi_g$ and $\phi_{a_i}$, to
\begin{align}
    D_\eta = \frac{\partial}{\partial \eta},\qquad
    D_F    = \frac{\partial}{\partial F},\qquad
    D_V    = \frac{\partial}{\partial V}.
\end{align}

Using the above in the determining equations \eqref{eq:AUX_D_x_eq_zero}, we get,
    \begin{align}\label{eq:CONT_final_phig_phiai_functionals}
        {\phi_g = \phi_g(g),} \qquad
        {\phi_{a_i} = \phi_{a_i}(g,R_\rho,\ell_0,C_p,\chi).}
    \end{align}

\subsection{ {Continuity equation}}
The linearized symmetry condition for \eqref{eq:RElanzaCont} is,
\begin{align}\label{eq:LSC_statement_cont}
X^{(2)}[\Delta_C] = 0,
\end{align}
The continuity equation is a first-order ODE, therefore the coefficients up to first order in
\(X^{(2)}\) enter the LSC, and the second order prolongation terms vanish identically. Additionally, the terms corresponding to arbitrary functions also drop out for this case. From \eqref{eq:LSC_statement_cont}, we have,
\begin{align}
    \phi^{(1)}_V + \phi_F = 0
\end{align}
Substituting $\phi^{(1)}_V=D_{\eta}(\phi_V)-V^{(1)}D_{\eta}(\phi_{\eta})$ and expanding the total derivative $D_\eta = \partial_\eta + F^{(1)}\partial_F + V^{(1)}\partial_V + g^{(1)}\partial_g$, we obtain,
\begin{align*}
  \partial_\eta \phi_V + F^{(1)} \partial_F \phi_V + V^{(1)} \partial_V \phi_V
     + g^{(1)} \partial_g \phi_V  -  V^{(1)}\bigl(
       \partial_\eta \phi_\eta + F^{(1)} \partial_F \phi_\eta \\ \quad 
       + V^{(1)} \partial_V \phi_\eta + g^{(1)} \partial_g \phi_\eta
     \bigr) + \phi_F=0 .
\end{align*}
Using $V^{(1)}=-F$, collecting the coefficients of the independent jet variables, and grouping the remaining zeroth--order terms, results in,
\begin{align*}
F^{(1)}\bigl(\partial_F \phi_V + F \partial_F \phi_\eta\bigr)
+ g^{(1)}\bigl(\partial_g \phi_V + F \partial_g \phi_\eta\bigr)
+ F\bigl(-\partial_V \phi_V + \partial_\eta \phi_\eta\bigr)
\\ \quad - F^2 \partial_V \phi_\eta
+ \partial_\eta \phi_V + \phi_F = 0.
\end{align*}
The aforementioned convention on superscripts applies here and throughout the main article. For example, $F^{(1)}:=\partial F/\partial \eta$ etc.\par 
Splitting on $F^{(1)},  g^{(1)}$ and the remaining zero-order terms, we get the determining equations for the equivalence symmetry analysis of the continuity equation as listed below,
\begin{subequations}\label{eq:cont_all_deteqs}
\begin{align}
\partial_F \phi_V + F \partial_F \phi_\eta &= 0,\label{eq:cont_det1} \\
\partial_g \phi_V + F \partial_g \phi_\eta &= 0,\label{eq:cont_det2} \\
 F\bigl(-\partial_V \phi_V + \partial_\eta \phi_\eta\bigr) - F^2 \partial_V \phi_\eta
+ \partial_\eta \phi_V + \phi_F &= 0.
\end{align}
\end{subequations}

To solve the above system to derive meaningful dependencies, we assume the spatial coordinate transformation does not depend on the velocities or temperature, implying $\phi_\eta = \phi_\eta(\eta)$, which readily yields,

\begin{subequations}\label{eq:cont_det_all2}
\begin{align}
\left.
\begin{aligned}
\partial_F \phi_V   &= 0, \\
\partial_F \phi_\eta &= 0,
\end{aligned}
\right\}
&\quad \text{from \eqref{eq:cont_det1}}
\label{eq:cont_det11}
\\[0.25em]
\left.
\begin{aligned}
\partial_g \phi_V   &= 0, \\
\partial_g \phi_\eta &= 0,
\end{aligned}
\right\}
&\quad \text{from \eqref{eq:cont_det2}}
\label{eq:cont_det22}
\end{align}
\end{subequations}
The first of both equations \eqref{eq:cont_det11} and \eqref{eq:cont_det22} give, 
\begin{align}\label{eq:cont_phiv_equals_phiv_of_V_eta}
    \phi_V = \phi_V(\eta, V)
\end{align}

Now using the above together with $\phi_\eta = \phi_\eta(\eta)$ in the zero-order split, we get,
    \begin{align}
       \phi_F= F \left( \partial_V \phi_V - \partial_\eta \phi_\eta \right)-\partial_\eta \phi_V  \label{eq:cont_forphiF_new}.
    \end{align}

\subsection{ {Momentum equation}}
In this section, the prolonged infinitesimal generator \eqref{eq:X2} is projected onto the momentum equation \eqref{eq:RElanzaMom} to yield the linearized symmetry condition as,
\begin{align}\label{eq:MOM_X2[DM]}
    X^{(2)}[\Delta_M] = 0
\end{align}
Since \eqref{eq:RElanzaMom} contains the second order derivative of $F$ and first order derivatives of $F \text{ and } g$ with respect to $\eta$, and of $\ell_0$ with respect to $g$, the prolongations corresponding to these terms will be non-zero. The reduced linearized symmetry condition \eqref{eq:MOM_X2[DM]} yields,
\begin{align}\label{eq:mom_LSC_terms}
    \phi_F^{(2)} ={}& \big( \phi_\eta \partial_\eta +\phi_V \partial_V + \phi_F \partial_F + \phi_g \partial_g + \phi_{\ell_0} \partial_{\ell_0} + \phi_{R_\rho} \partial_{R_\rho} \notag \\ & + \phi^{(1)}_F \partial_{F^{(1)}} + \phi^{(1)}_g \partial_{g^{(1)}} + \phi^{(1)g}_{\ell_0} \partial_{\ell_0^{(g)}} \big)  \mathbb{M},
\end{align}
where the following shorthand has been adopted,
\begin{align*}
    \mathbb{M}=\frac{1}{l_0} \left[ V\frac{dF}{d\eta} - \frac{1}{2} (R_\rho - F^{2}) -\frac{dF}{d\eta} \frac{dl_0}{dg} \frac{dg}{d\eta} \right]
\end{align*}
Before we expand \eqref{eq:mom_LSC_terms}, the prolonged infinitesimals $\phi_g^{(1)}, \phi_F^{(1)}, \phi^{(1)g}_{\ell_0} $ and $\phi_F^{(2)}$ can be obtained as follows,
\begin{subequations}\label{eq:MOM_phiF1_phiF2}
    \begin{align}
        \phi_g^{(1)} &= D_\eta \big ( \phi_g \big ) - \frac{dg}{d\eta} D_\eta \big ( \phi_\eta \big ) \label{eq:MOM_phig(1)TotalDerivativeForm} \\
        \phi_F^{(1)} &= D_\eta \big ( \phi_F \big ) - \frac{dF}{d\eta} D_\eta \big ( \phi_\eta \big ) \\
        \phi_F^{(2)} &= D_\eta \big ( \phi^{(1)}_F \big ) - \frac{d^2 F}{d\eta^2} D_\eta \big ( \phi_\eta \big ) 
    \end{align}
\end{subequations}
 \par 
Using $D_\eta = \partial_\eta +(dF/d\eta)\partial_F +(dV/d\eta)\partial_V + (dg/d\eta)\partial_g + d^2F/d\eta^2 \partial_{F^{(1)}} + d^2g/d\eta^2 \partial_{g^{(1)}} + \ldots$ together with $\phi_g=\phi_g(g)$ from \eqref{eq:CONT_final_phig_phiai_functionals}, $\phi_\eta=\phi_\eta(\eta)$ and $\phi_F$ from \eqref{eq:cont_forphiF_new}, we get, 
\begin{subequations}\label{eq:MOM_phig1_phiF1_phiF2_final_forms}
    \begin{align}
        \phi_g^{(1)} &= \left( \frac{d}{dg} \big( \phi_g(g) \big)  - \frac{d}{d\eta} \big( \phi_\eta(\eta) \big) \right) \frac{dg}{d\eta}, \label{eq:PHI1G_MOM_SPLIT____}\\
        \phi_F^{(1)} &= -F^2 \partial_{VV} \phi_V - F \frac{d^2}{d\eta^2}\big(\phi_\eta(\eta)\big)+2F\partial_{V\eta}\phi_V + \frac{dF}{d \eta}\partial_V \phi_V - 2 \frac{dF}{d\eta} \frac{d}{d\eta} \big( \phi_\eta(\eta) \big) - \partial_{\eta \eta} \phi_V , \\
        \phi_F^{(2)} &= \mathbb{M} \left( \partial_V \phi_V - 3\frac{d}{d\eta} \big( \phi_\eta(\eta) \big) \right) -3 F \frac{dF}{d\eta}\partial_{VV}\phi_V + 3 \frac{dF}{d\eta}\partial_{\eta V}\phi_V - 3 \frac{dF}{d\eta} \frac{d^2\big( \phi_\eta(\eta) \big)}{d\eta^2} \notag \\ &\quad {} + F^3\partial_{VVV}\phi_V - 3F^2 \partial_{\eta V V} \phi_V -  F\frac{d^3\big( \phi_\eta(\eta) \big)}{d\eta^3} + 3F\partial_{\eta \eta V} \phi_V - \partial_{\eta \eta \eta} \phi_V
    \end{align}
\end{subequations}
\par
Evaluating the differentials on the right-hand side of \eqref{eq:mom_LSC_terms} termwise, we obtain,
\begin{equation}
\left.
\begin{aligned}
\partial_\eta \mathbb{M} &= 0, \;
\partial_V \mathbb{M} = \frac{1}{\ell_0} \frac{dF}{d\eta}, \;
\partial_F \mathbb{M} = \frac{F}{\ell_0}, \;
\partial_g \mathbb{M} = 0, \;
\partial_{\ell_0} \mathbb{M} = -\frac{1}{\ell_0} \frac{d^{2}F}{d\eta^{2}}, \;
\partial_{R_\rho} \mathbb{M} = -\frac{1}{2\ell_0}, \\[0.3em]
\partial_{F^{(1)}} \mathbb{M}
  &= \frac{1}{\ell_0}\!\left( V - \frac{d\ell_0}{dg} \frac{dg}{d\eta} \right), \;
\partial_{g^{(1)}} \mathbb{M}
  = -\frac{1}{\ell_0}\!\left( \frac{d\ell_0}{dg} \frac{dF}{d\eta} \right), \;
\partial_{\ell_0^{(g)}} \mathbb{M}
  = -\frac{1}{\ell_0} \frac{dg}{d\eta} \frac{dF}{d\eta}
\end{aligned}
\right\}
\label{eq:MOM_termwise_partials_of_M}
\end{equation}
Substituting \eqref{eq:MOM_phig1_phiF1_phiF2_final_forms} and \eqref{eq:MOM_termwise_partials_of_M} in \eqref{eq:mom_LSC_terms}, and rearranging, we finally obtain the expanded linearized symmetry condition,

\begin{align}\label{eq:MOM_expandedLSC_of_mom}
&\frac{dF}{d\eta}\frac{dg}{d\eta}
\left[
    -\frac{1}{\ell_0}\frac{d\ell_0}{dg}\phi_{\ell_0}
    + \phi_{\ell_0}^{(1)g}
    + \frac{d\ell_0}{dg}\frac{d}{dg}\big(\phi_g(g)\big)
\right]
\notag\\
&\quad
+ \frac{dF}{d\eta}
\left[
    -\phi_V
    + 3\ell_0 \partial_{\eta V}\phi_V
    - 3\ell_0 F \partial_{VV}\phi_V
    - 3\ell_0 \frac{d^2}{d\eta^2}\big(\phi_\eta(\eta)\big)
    - V\frac{d}{d\eta}\big(\phi_\eta(\eta)\big)
    + \frac{\phi_{\ell_0}}{\ell_0}V
\right]
\notag\\
&\quad
+ \frac{dg}{d\eta}
\left[
    -\frac{d\ell_0}{dg}
    \left(
        \partial_{\eta\eta}\phi_V
        - 2F\partial_{\eta V}\phi_V
        + F^2\partial_{VV}\phi_V
        + F\frac{d^2}{d\eta^2}\big(\phi_\eta(\eta)\big)
    \right)
\right]
\notag\\
&\quad
+
\left[
    -\ell_0\partial_{\eta\eta\eta}\phi_V
    + F\partial_\eta\phi_V
    + 3\ell_0 F\partial_{\eta\eta V}\phi_V
    - \frac{1}{2}F^2\partial_V\phi_V
    - 3\ell_0 F^2\partial_{\eta VV}\phi_V
\right.
\notag\\
&\qquad\qquad
    + \ell_0 F^3\partial_{VVV}\phi_V
    - \frac{1}{2}F^2\frac{d}{d\eta}\big(\phi_\eta(\eta)\big)
    - \ell_0 F\frac{d^3}{d\eta^3}\big(\phi_\eta(\eta)\big)
\notag\\
&\qquad\qquad
    + \frac{F^2}{2\ell_0}\phi_{\ell_0}
    + \frac{1}{2}\phi_{R_\rho}
    - \frac{1}{2}R_\rho\partial_V\phi_V
    + \frac{3}{2}R_\rho\frac{d}{d\eta}\big(\phi_\eta(\eta)\big)
\notag\\
&\qquad\qquad\left.
    - \frac{R_\rho}{2\ell_0}\phi_{\ell_0}
    + V\partial_{\eta\eta}\phi_V
    - 2FV\partial_{\eta V}\phi_V
    + F^2V\partial_{VV}\phi_V
    + FV\frac{d^2}{d\eta^2}\big(\phi_\eta(\eta)\big)
\right]
=0.
\end{align}
\par 
We have eliminated the highest derivative by substituting $F_{\eta\eta}=\mathbb{M}$ in \eqref{eq:MOM_expandedLSC_of_mom}. The first derivatives \(dF/d\eta\) and \(dg/d\eta\) are independent jet
coordinates. Therefore, the coefficients of the independent monomials $\frac{dF}{d\eta}\frac{dg}{d\eta}, \frac{dF}{d\eta}, \frac{dg}{d\eta}$ and the remaining constant terms must vanish separately. This gives the following determining equations,

\begin{align}
&\text{ coeff. of }\frac{dF}{d\eta}\frac{dg}{d\eta}: \notag \\
&-\frac{1}{\ell_0}\frac{d\ell_0}{dg}\phi_{\ell_0}
+ \phi_{\ell_0}^{(1)g}
+ \frac{d\ell_0}{dg}
\frac{d}{dg}\big(\phi_g(g)\big)
=0, 
\label{eq:mom_split_Fg}
\\[1ex]
&\text{ coeff. of }\frac{dF}{d\eta}: \notag \\
&-\phi_V
+3\ell_0\partial_{\eta V}\phi_V
-3\ell_0F\partial_{VV}\phi_V
-3\ell_0\frac{d^2}{d\eta^2}\big(\phi_\eta(\eta)\big)
-V\frac{d}{d\eta}\big(\phi_\eta(\eta)\big)
+\frac{\phi_{\ell_0}}{\ell_0}V
=0,
\label{eq:mom_split_F}
\\[1ex]
&\text{ coeff. of }\frac{dg}{d\eta}: \notag \\
&-\frac{d\ell_0}{dg}
\left(
    \partial_{\eta\eta}\phi_V
    -2F\partial_{\eta V}\phi_V
    +F^2\partial_{VV}\phi_V
    +F\frac{d^2}{d\eta^2}\big(\phi_\eta(\eta)\big)
\right)
=0,
\label{eq:mom_split_g}
\\[1ex]
&\text{ remaining zero-order terms}: \notag \\
&-\ell_0\partial_{\eta\eta\eta}\phi_V
+F\partial_\eta\phi_V
+3\ell_0F\partial_{\eta\eta V}\phi_V
-\frac{1}{2}F^2\partial_V\phi_V
-3\ell_0F^2\partial_{\eta VV}\phi_V
+\ell_0F^3\partial_{VVV}\phi_V
\notag\\
&\quad
-\frac{1}{2}F^2\frac{d}{d\eta}\big(\phi_\eta(\eta)\big)
-\ell_0F\frac{d^3}{d\eta^3}\big(\phi_\eta(\eta)\big)
+\frac{F^2}{2\ell_0}\phi_{\ell_0}
+\frac{1}{2}\phi_{R_\rho}
-\frac{1}{2}R_\rho\partial_V\phi_V
\notag\\
&\quad
+\frac{3}{2}R_\rho\frac{d}{d\eta}\big(\phi_\eta(\eta)\big)
-\frac{R_\rho}{2\ell_0}\phi_{\ell_0}
+V\partial_{\eta\eta}\phi_V
-2FV\partial_{\eta V}\phi_V
+F^2V\partial_{VV}\phi_V
+FV\frac{d^2}{d\eta^2}\big(\phi_\eta(\eta)\big)
=0.
\label{eq:mom_split_zero}
\end{align}

We now solve these determining equations one by one. Equation \eqref{eq:mom_split_Fg} yields, 

\begin{align}
    \phi^{(1)g}_{\ell_0} = \left( \frac{\phi_{\ell_0}}{\ell_0} - \frac{d}{dg}\big( \phi_g(g) \big) \right) \frac{d\ell_0}{dg}.
\end{align}

Splitting \eqref{eq:mom_split_F} by $F$, we obtain,
\begin{align}\label{eq:phiV_equal_AVPLUSB____}
    3\ell_0 \partial_{VV}\phi_V = 0 \Rightarrow \phi_V = \mathcal{A}(\eta)V+ \mathcal{B}(\eta),
\end{align}
which immediately gives $\partial_{\eta V}\phi_V = \partial_\eta \mathcal{A}(\eta)$. Using these in \eqref{eq:mom_split_F}, we get, 

\begin{align}
    -(\mathcal{A}(\eta)V+\mathcal{B}(\eta)) + 3 \ell_0 \frac{d\mathcal{A}(\eta)}{d \eta} - 3 \ell_0 \frac{d^2}{d \eta^2} \big( \phi_\eta(\eta) \big) - V \frac{d}{d \eta} \big( \phi_\eta(\eta) \big) +  {V}\frac{\phi_{\ell_0}}{\ell_0} =0
\end{align}

Splitting by the powers of $V$ we get two equations, one for $V^1$ and the other for $V^0$, respectively, as,
\begin{subequations}
    \begin{align}
    &\phi_{\ell_0} = \left( \mathcal{A}(\eta) + \frac{d}{d \eta} \big( \phi_\eta(\eta) \big)  \right) \ell_0, \\
    -&\mathcal{B}(\eta) + 3\ell_0 \left( \frac{d\mathcal{A}(\eta)}{d \eta} - \frac{d^2}{d \eta^2} \big( \phi_\eta(\eta) \big)  \right)=0 \Rightarrow \mathcal{B}(\eta) = 0, \text{ and } \frac{d\mathcal{A}(\eta)}{d \eta} =  \frac{d^2}{d \eta^2} \big( \phi_\eta(\eta) \big)  \label{eq:Bequal0__mom_split__}
\end{align}
\end{subequations}

Similarly, splitting \eqref{eq:mom_split_g} by the powers of $F$, and skipping the intermediate steps for brevity, we obtain,   
\begin{subequations}
    \begin{align}
    \partial_{VV}\phi_V &= 0, \text{ for the coeff. of } F^2 \\
    \frac{d^2}{d \eta^2} \big( \phi_\eta(\eta) \big) &= 2 \partial_{\eta V}\phi_V, \text{ for the coeff. of } F^1 \label{eq:F1-split_MOM__}\\
    \partial_{\eta \eta} \phi_V &= 0, \text{ for the coeff. of } F^0
\end{align}
\end{subequations}

Using $\mathcal{B}(\eta)=0$ from \eqref{eq:Bequal0__mom_split__} in \eqref{eq:phiV_equal_AVPLUSB____}, we get $\phi_V = \mathcal{A}(\eta) V$, which leads to $\partial_{\eta V} \phi_V = d\mathcal{A}(\eta)/d\eta$. With this and \eqref{eq:F1-split_MOM__}, we get, 
\begin{align}
     \frac{d^2}{d \eta^2} \big( \phi_\eta(\eta) \big) = 2 \frac{d\mathcal
     A(\eta)}{d\eta}
\end{align}
The above equation and the last of \eqref{eq:Bequal0__mom_split__} are satisfied only when, 
\begin{align}
    \frac{d\mathcal{A}(\eta)}{d\eta} = 0 \Rightarrow \mathcal{A}(\eta) = A_0, \text{a constant}
\end{align}
which immediately yields, 
\begin{align}
    &\frac{d^2}{d \eta^2} \big( \phi_\eta(\eta) \big) = 0 \Rightarrow \frac{d}{d \eta} \big( \phi_\eta(\eta) \big) = C_0, \text{ a constant }\\
    &\text{such that}, \  \phi_\eta = C_0 \eta 
\end{align}
where the constant of integration is chosen to be zero to preserve the transformed wall coordinate at $\hat \eta = 0$. 
\par 
Substituting the results obtained so far in \eqref{eq:mom_split_zero} and \eqref{eq:cont_forphiF_new}, we obtain,  
\begin{align}
    \phi_{R_\rho} &= 2(A_0-C_0)R_\rho, \\
    \phi_F &= (A_0 - C_0) F,
\end{align}
respectively. 
\par 
The infinitesimals generated so far by the preceding analysis are listed below,
\begin{equation}\label{eq:SUMMARY_OF_SYMMS_MOM__}
\left.
    \begin{aligned}
    \phi_\eta = C_0 \eta, \quad &\phi_V = A_0 V, \quad \phi_F = (A_0 - C_0) F, \quad \phi_g = \phi_g(g),\\ &\phi_{R_\rho} = 2(A_0-C_0)R_\rho,
    \quad \phi_{\ell_0} = (A_0 + C_0 )\ell_0.
\end{aligned}
\right\}
\end{equation}

\subsection{ {Energy equation}}
Following the same procedure as above, the prolonged infinitesimal generator \eqref{eq:X2} is projected onto \eqref{eq:RElanzaEnergy} to yield the linearized symmetry condition for the energy equation as,
\begin{align}\label{eq:ENERGY_X2[DE]}
    X^{(2)}[\Delta_E] = 0
\end{align}

The energy equation \eqref{eq:RElanzaEnergy} contains the first and second order derivatives of $g$ with respect to $\eta$, and first order derivative of $\chi$ with respect to $g$. Therefore, the prolongations corresponding to these terms will be non-zero. The reduced form of the linearized symmetry condition \eqref{eq:ENERGY_X2[DE]} reads,
\begin{align}\label{eq:energy_LSC_terms}
    \phi_g^{(2)} ={}& \big( \phi_\eta \partial_\eta +\phi_V \partial_V + \phi_g \partial_g + \phi_{\chi} \partial_{\chi} + \phi_{C_p} \partial_{C_p} + \phi^{(1)}_g \partial_{g^{(1)}} + \phi^{(1)g}_{\chi} \partial_{\chi^{(g)}} \big)  \mathbb{E},
\end{align}
where the following shorthand has been adopted,
\begin{align*}
    \mathbb{E}=\frac{1}{\chi} \left[ C_p V\frac{dg}{d\eta} - \left ( \frac{dg}{d\eta} \right)^2 \frac{d\chi}{dg} \right].
\end{align*}
From \eqref{eq:PHI1G_MOM_SPLIT____} and \eqref{eq:SUMMARY_OF_SYMMS_MOM__}, the first prolongation $\phi^{(1)}_g = ({d\phi_g(g)}/{dg} -C_0)dg/d\eta $. Using this, we obtain the second prolongation $\phi^{(2)}_g$ recursively as,
\begin{align}\label{eq:ENERGY_phig(2)}
    \phi^{(2)}_g &= D_\eta \big(\phi^{(1)}_g \big) - \frac{d^2g}{d\eta^2} D_\eta \big( \phi_\eta \big) \notag \\
    \Rightarrow \phi^{(2)}_g &= \left( \frac{dg}{d\eta} \right)^2 \frac{d^2 }{dg^2}\big( \phi_g(g) \big)  + \left( \frac{d\phi_g(g)}{dg} -2C_0 \right) \frac{d^2g}{d\eta^2}
\end{align}
\par 
Evaluating the differentials on the right-hand side of \eqref{eq:energy_LSC_terms} termwise, we obtain,
\begin{equation}
\left.
\begin{aligned}
\partial_\eta \mathbb{E} &= 0, \; \;
\partial_V \mathbb{E} = C_p\frac{1}{\chi} \frac{dg}{d\eta}, \; \;
\partial_g \mathbb{E} = 0, \;
\partial_{\chi} \mathbb{E} = -\frac{1}{\chi} \frac{d^{2}g}{d\eta^{2}}, \; \\[0.3em]
\partial_{C_p} \mathbb{E} &= \frac{1}{\chi}V\frac{dg}{d\eta}, \; \;
\partial_{g^{(1)}} \mathbb{E}
  = \frac{1}{\chi}\!\left( C_p V - 2 \frac{d\chi}{dg} \frac{dg}{d\eta} \right), \; \;
\partial_{\chi^{(g)}} \mathbb{E}
  = -\frac{1}{\chi} \left( \frac{dg}{d\eta} \right)^2
\end{aligned}
\right\}
\label{eq:ENERGY_termwise_partials_of_E}
\end{equation}
Using $\phi^{(1)}_g, \phi^{(2)}_g$ and \eqref{eq:ENERGY_termwise_partials_of_E} in \eqref{eq:energy_LSC_terms}, and upon simplifying and rearranging, we finally obtain the expanded linearized symmetry condition,
\begin{align}\label{eq:ENERGY_expandedLSC_of_energy}
    &V\frac{dg}{d\eta}
    \left[
        -(A_0 + C_0)C_p
        + C_p \frac{\phi_\chi}{\chi}
        -\phi_{C_p}
    \right]
    \notag\\
    &\quad
    + \left(\frac{dg}{d\eta}\right)^2
    \left[
        \chi \frac{d^2}{dg^2}\big(\phi_g(g)\big)
        + \frac{d\chi}{dg}\frac{d}{dg}\big(\phi_g(g)\big)
        + \phi^{(1)g}_\chi
        - \frac{d\chi}{dg}\frac{\phi_\chi}{\chi}
    \right]
    =0 .
\end{align}

Splitting the above equation with respect to the independent monomials $Vdg/d\eta$ and $(dg/d\eta)^2$, we obtain the following, respectively,
\begin{align}\label{eq:ENERGY_expression_of_phiCp}
    \phi_{C_p} &=  \left( -(A_0 + C_0) + \frac{\phi_\chi}{\chi} \right)C_p \\
    \chi \frac{d^2}{dg^2} \big( \phi_g(g) \big) &=  \frac{\phi_\chi}{\chi} \frac{d\chi}{dg} - \phi^{(1)g}_\chi - \frac{d\chi}{dg} \frac{d}{dg} \big(\phi_g(g) \big)
\end{align}
Using $\phi^{(1)g}_{\chi} = D_g \big( \phi_\chi \big)- (d\chi/dg)D_g\big( \phi_g (g)\big) = D_g \big( \phi_\chi \big)- (d\chi/dg)(d\phi_g(g)/dg)$ in the second equation above yields,
\begin{align}\label{eq:ENERGY_expression_of_phiChi}
    \chi \frac{d^2}{dg^2} \big( \phi_g(g) \big)= \frac{\phi_\chi}{\chi} \frac{d\chi}{dg} -D_g \big(\phi_\chi \big) 
\end{align}
Equations \eqref{eq:ENERGY_expression_of_phiCp} and \eqref{eq:ENERGY_expression_of_phiChi} must hold for all admissible \(C_p(g)\) and \(\chi(g)\). Recall, from \eqref{eq:CONT_final_phig_phiai_functionals}, we have established that $\phi_{\chi} = \phi_{\chi}(g,R_\rho,\ell_0,C_p,\chi)$, therefore we can expand $D_g$ as,
\begin{align}
    D_g(\phi_\chi)
=
\frac{\partial \phi_\chi}{\partial_g}
+
\frac{\partial R_\rho}{\partial g} \frac{\partial \phi_\chi}{\partial R_\rho}
+
\frac{\partial \ell_0}{\partial g} \frac{\partial \phi_\chi}{\partial \ell_0}
+
\frac{\partial C_p}{\partial g} \frac{\partial \phi_\chi}{\partial C_p}
+
\frac{\partial \chi}{\partial g} \frac{\partial \phi_\chi}{\partial \chi}
\end{align}
Substituting this into \eqref{eq:ENERGY_expression_of_phiChi} gives,
\begin{align}
\chi \,\frac{d^{2}}{dg^{2}}(\phi_{g}(g))
=
\frac{d\chi}{dg}
\left(
\frac{\phi_{\chi}}{\chi}
-
\frac{\partial \phi_{\chi}}{\partial \chi}
\right)
-
\frac{\partial \phi_{\chi}}{\partial g}
-
\frac{dR_{\rho}}{dg}\,\frac{\partial \phi_{\chi}}{\partial R_{\rho}}
-
\frac{d\ell_{0}}{dg}\,\frac{\partial \phi_{\chi}}{\partial \ell_{0}}
-
\frac{dC_{p}}{dg}\,\frac{\partial \phi_{\chi}}{\partial C_{p}}.
\end{align}
The left-hand side is a function of \(g\) only, since
$\phi_{g}=\phi_{g}(g)$ from \eqref{eq:CONT_final_phig_phiai_functionals}.

Since $R_\rho(g), \ell_0(g), C_p(g)$, and $\chi(g)$ are arbitrary elements of the equivalence class, their first derivatives with respect to \(g\) are treated as independent auxiliary jet variables. Therefore, the above identity must hold for arbitrary values of  ${dR_{\rho}}/{dg},{d\ell_{0}}/{dg}, {dC_{p}}/{dg} \text{ and } {d\chi}/{dg}$ and the coefficients multiplying these independent quantities are split and set
to zero separately, together with the remaining derivative-free part,
\begin{subequations}\label{eq:phi_chi_split_coeffs}
\begin{align}
\text{coefficient of } \frac{dR_{\rho}}{dg}
\qquad &\Rightarrow \qquad
\frac{\partial \phi_{\chi}}{\partial R_{\rho}}=0,
\label{eq:phi_chi_split_coeffs_a1}\\[0.5em]
\text{coefficient of } \frac{d\ell_{0}}{dg}
\qquad &\Rightarrow \qquad
\frac{\partial \phi_{\chi}}{\partial \ell_{0}}=0,
\label{eq:phi_chi_split_coeffs_b2}\\[0.5em]
\text{coefficient of } \frac{dC_{p}}{dg}
\qquad &\Rightarrow \qquad
\frac{\partial \phi_{\chi}}{\partial C_{p}}=0,
\label{eq:phi_chi_split_coeffs_c3}\\[0.5em]
\text{coefficient of } \frac{d\chi}{dg}
\qquad &\Rightarrow \qquad \frac{\phi_{\chi}}{\chi} - \frac{\partial \phi_{\chi}}{\partial \chi}=0. \label{eq:phi_chi_split_coeffs_d4}
\\[0.5em]
\text{remaining terms, } \chi \,\frac{d^{2}}{dg^{2}}(\phi_{g}(g)) &= -\frac{\partial \phi_{\chi}}{\partial g} \label{eq:phi_chi_split_coeffs_e5}
\end{align}
\end{subequations}
Equation \eqref{eq:phi_chi_split_coeffs_d4} has general solution
\begin{align}\label{eq:energy_ansatz_phichi_fg}
    \phi_\chi=f(g)\chi,
\end{align}
$f(g)$ represents a $g$-dependent rescaling factor for $\chi$. Using \eqref{eq:energy_ansatz_phichi_fg}  in \eqref{eq:phi_chi_split_coeffs_e5} yields,
\begin{subequations}
    \begin{align}
    {}&\chi \left ( \frac{d^2}{dg^2} \big( \phi_g(g) \big) + \frac{d}{dg} \big( f(g)\big)  \right ) = 0, \notag \\
    {}&\Rightarrow f(g) = K_0 - \frac{d}{dg} \big( \phi_g (g)\big),
\end{align}
\end{subequations}
where $K_0$ is a constant of integration. 
\par 
Using $f(g)$ in \eqref{eq:ENERGY_expression_of_phiCp} and \eqref{eq:energy_ansatz_phichi_fg} yields,
\begin{subequations}
    \begin{align}
    \phi_{C_p} &= \left(-(A_0+C_0)+K_0 - \frac{d}{dg}\big( \phi_g (g)\big)  \right)C_p, \label{eq:ENERGY_final_phiCPP_expression} \\
    \phi_\chi &= \left( K_0 - \frac{d}{dg} \big( \phi_g (g)\big) \right) \chi. \label{eq:ENERGY_final_phiChi_expression}
\end{align}
\end{subequations}

Combining all the results, the infinitesimal generator can be written as,
\begin{align}
X
&=
C_0\eta\frac{\partial}{\partial \eta}
+
(A_0-C_0)F\frac{\partial}{\partial F}
+
A_0V\frac{\partial}{\partial V}
+
\phi_g(g)\frac{\partial}{\partial g}
+
2(A_0-C_0)R_\rho\frac{\partial}{\partial R_\rho}
+
(A_0+C_0)\ell_0\frac{\partial}{\partial \ell_0}
\notag\\
&\quad
+
\left[
K_0-\frac{d}{dg}\left(\phi_g(g)\right)
\right]\chi\frac{\partial}{\partial \chi}
+
\left[
K_0-(A_0+C_0)-\frac{d}{dg}\left(\phi_g(g)\right)
\right]C_p\frac{\partial}{\partial C_p}.
\end{align}

As a sanity check, direct substitution of the finite transformations for the arbitrary elements and other variables into the continuity, momentum, and energy equations shows that each equation is mapped onto itself, proving that the derived symmetries are correct and valid. The detailed equation-by-equation substitution is given in Appendix \ref{APPEND_INV_CHECK}.
} 



\section{{$g$-mapping and numerical verification }}\label{sec:gmpa_numver}
This section presents the key theoretical result in this paper: an integral invariant built from thermochemical property data, which is inverted numerically to infer the mapped non-dimensional temperature profile across \textcolor{blue}{the similarity reduced stagnation-point} boundary layer. Having obtained the expression for $\phi_\chi$ in
\eqref{eq:ENERGY_final_phiChi_expression}, which depends on $\chi$ and the infinitesimal of $g$, we proceed to infer the $g$-map when property tables for $\chi(g)$ are available for the target solution to which $g$ is to be mapped. In hypersonic boundary layers, the required thermochemical properties
(viscosity, thermal conductivity, specific heats, etc.) can be obtained
up to temperatures of order $T \sim 20{,}000~\mathrm{K}$ from
high-temperature libraries such as \textsc{Mutation++} \citep{mutation++ref}. {\color{magenta}The availability of the property law data in hypersonics motivates us to cast the symmetry relations in terms of tabulated thermochemical property data rather than prescribed analytic constitutive laws. The tabulated property data define a particular numerical representation of the arbitrary
elements. \citet{tranchant2025generalizing} provide a useful precedent for the present use of tabulated property laws. In one of their cases, they impose the tabulated opacity data in the radiative wave model in astrophysics to constrain the temperature map. The present construction as outlined below is analogous: we use tabulated $\chi$ law on both the flight and laboratory sides to constrain the temperature map $\hat{g}$. The rest of the arbitrary elements are also constrained by this reparametrization, and their deviations with respect to the flight values serve as the indicators of how the true flight data compare against the transformed counterparts. All of this is discussed in the rest of this section.}
\par 
We regard the one–parameter equivalence group as
generating an \emph{orbit} in the extended space of variables. The tangent vector to the orbit at the point $(\hat{g},\hat{\chi})$ is $ {d\hat{g}}/{d\varepsilon}
    = \phi_g(\hat{g}),\
  {d\hat{\chi}}/{d\varepsilon}
    = \phi_\chi(\hat{g},\hat{\chi})$\citep{hydon2000}.
These equations describe the evolution of the transformed variables $(\hat{g},\hat{\chi})$ along the orbit, where the rate of change at each step is given by the infinitesimals $\phi_g \text{ and } \phi_\chi$ evaluated at that point $(\hat{g},\hat{\chi})$. To infer the finite map point by point in the boundary layer, we then consider the associated family of orbits through each initial point $g_0$ on the BL profile, where $ \hat g(\varepsilon = 0,g_0)=g_0$.
\par
Defining the Jacobian of this transformation by,
\begin{equation} J(\varepsilon,g_0) = \frac{\partial \hat g(\varepsilon,g_0)}{\partial g_0}. 
\end{equation}

Then, differentiating $J$ with respect to $\varepsilon$ and interchanging the order of differentiation gives,
\begin{equation} \frac{\partial J}{\partial \varepsilon} = \frac{\partial}{\partial \varepsilon} \left( \frac{\partial \hat g}{\partial g_0} \right) = \frac{\partial}{\partial g_0} \left( \frac{\partial \hat g}{\partial \varepsilon} \right). 
\end{equation}

Using $\partial \hat{g}/\partial\varepsilon=\phi_g(\hat g)$, this gives, \begin{align} 
\frac{\partial J}{\partial \varepsilon} = \frac{\partial}{\partial g_0}\phi_g(\hat{g}(\varepsilon,g_0)) = \frac{d}{d\hat{g}(\varepsilon,g_0)} \big(\phi_g(\hat{g}(\varepsilon,g_0)) \big) \frac{\partial \hat g}{\partial g_0} =\frac{d}{d\hat{g}(\varepsilon,g_0)} \big(\phi_g(\hat{g}(\varepsilon,g_0)) \big) J
\end{align}
\par 
Integrating the above equation yields,
\begin{equation}\label{eq:INFER_Jeps_expression_final_RE}
J(\varepsilon,g_0) = \exp\left( \int_0^\varepsilon \frac{d}{d\hat{g}(s,g_0)} \big(\phi_g(\hat{g}(s,g_0)) \big)\,ds \right).
\end{equation}
where $J(\varepsilon=0,g_0)=1$ sets the constant of integration as $\ln\big[J(0,g_0)\big]=0$.  

Using \eqref{eq:ENERGY_final_phiChi_expression}, the transformation of $\chi$ satisfies,
\begin{align}
    \frac{\partial \hat{\chi}(\hat{g}(\varepsilon,g_0))}{\partial  \varepsilon} = \left( K_0 -  \frac{d}{d\hat{g}(\varepsilon,g_0)} \big( \phi_g (\hat{g}(\varepsilon,g_0))\big) \right) \hat{\chi}(\hat{g}(\varepsilon,g_0)). 
\end{align}

Upon integration and using \eqref{eq:INFER_Jeps_expression_final_RE}, we obtain,
\begin{align}\label{eq:CHI_J_LAW}
    \hat{\chi}(\hat{g}(\varepsilon,g_0)) = \hat{\chi}(\hat{g}(0,g_0)) e^{K_0\varepsilon} \big( J(\varepsilon,g_0)\big) ^{-1}  
\end{align}
which is a pointwise differential identity for each point $g_0$ and its transformed state $\hat{g}$ in the boundary layer. {An analogous expression for $C_p$ can be immediately obtained as $\hat C_p(\hat g) = C_p(g)e^{(K_0-(A_0+C_0))\varepsilon}J^{-1}$. }
Invoking the definition of $J(\varepsilon,g_0)= {\partial \hat{g}(\varepsilon,g_0) }/{\partial g_0}$, and integrating \eqref{eq:CHI_J_LAW} both sides from any interior point $g_\text{interior}$ to the edge $g=1$, we finally get,
\begin{align}\label{eq:INTEGRAL_INVARIANT_SEC3}
     \int_{\hat{g}_\text{interior}}^1  {\chi}_{_{\varepsilon}}(\hat{g}) {d\hat{g} } = e^{K_0\varepsilon} \int_{g_\text{interior}}^1  \chi_{_{0}}(g) {d g} 
\end{align}
where at $\varepsilon=0,\  \hat{g}(0,g_0)=g_0$. The shorthand $\hat{\chi}(\hat{g}(\varepsilon,g_0))=\chi_{_{\varepsilon}}(\hat{g})$, $\hat{\chi}(\hat{g}(0,g_0))=\chi_{_{0}}(g_0)$ has been adopted and the dummy variable $g_0$ has been relabeled as $g$ in the right-hand side for simplicity. This is the integral invariant which connects the laboratory- and flight-side thermochemcial property data ($\chi$) that can be inverted numerically to infer the correct mapped $\hat{g}$, as demonstrated below.
\par
\begin{figure}[htbp]
     \centering
     \begin{subfigure}[b]{0.32\textwidth}
         \centering
         \includegraphics[width=0.8\textwidth]{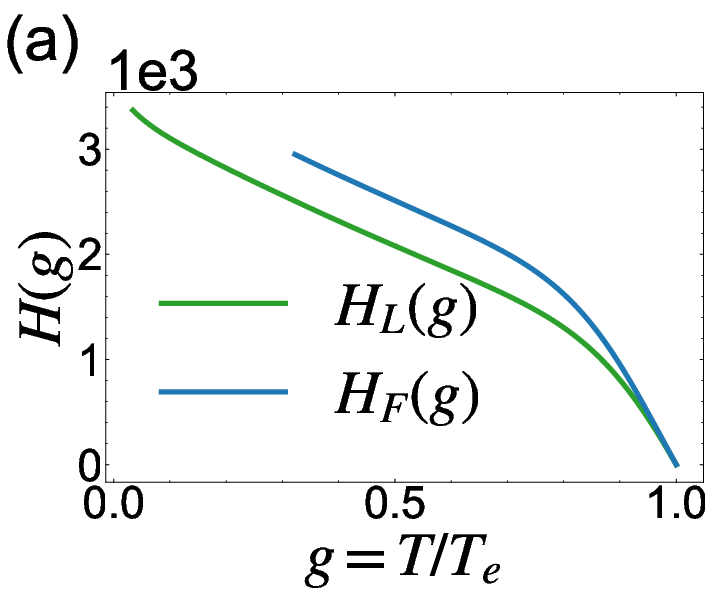}
         \label{fig:1Hplot1}
     \end{subfigure}
     \hfill
     \begin{subfigure}[b]{0.32\textwidth}
         \centering
         \includegraphics[width=0.8\textwidth]{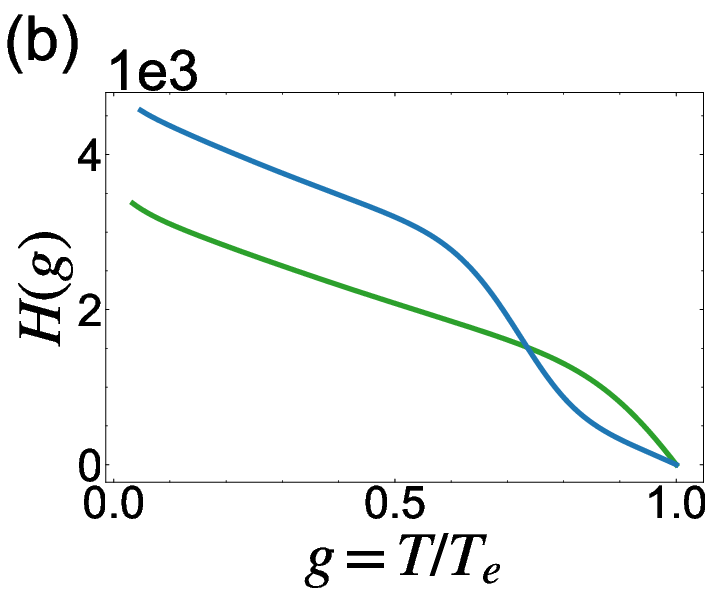}
         \label{fig:2Hplot2}
     \end{subfigure}
     \hfill
     \begin{subfigure}[b]{0.32\textwidth}
         \centering
         \includegraphics[width=0.8\textwidth]{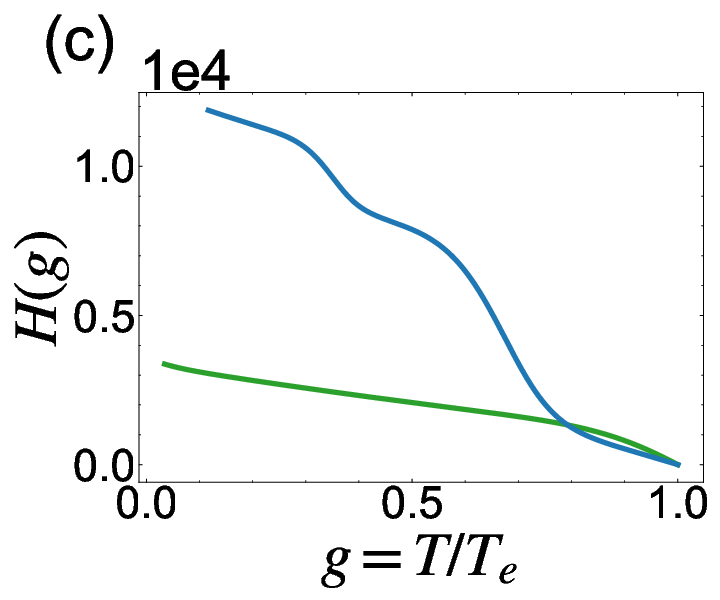}
         \label{fig:3Hplot3}
     \end{subfigure} \vspace{0mm}
        \caption{Integrals $H(g)$ for (a) $T_L^w=100$ K, $T_F^w=1000$ K and $T_L^e=2935$ K, $T_F^e=3103$ K, (b) $T_L^w=100$ K, $T_F^w=200$ K and $T_L^e=2935$ K, $T_F^e=4165$ K, and (c) $T_L^w=100$ K, $T_F^w=1000$ K and $T_L^e=2935$ K, $T_F^e=8586$ K. The integrals are monotone with respect to $g$. The abscissa $g$ is an arbitrary uniform g-grid selected to compute the integrals with the given $T^w, T^e$ for both laboratory and flight cases.}
        \label{fig:all_H_pots}
\end{figure}

 
Absorbing the parameter $\varepsilon$ in the constant, and defining $H_F(\hat{g}) = \int_{\hat{g}_\text{interior}}^1  \chi_{_{\varepsilon}}({s}) {d{s} }  $ and $H_L(g) = \int_{g_\text{interior}}^1  \chi_{_{0}}(s) ds  $, the finite transformation to obtain the mapped $\hat{g}$ is, 
\vspace{0mm}
\begin{align}\label{eq:INFER_FINAL_HF_HL_TRANSFORMATION}
\hat{g} &= H_F^{-1} \big( e^{K_0} H_L(g) \big ),
\end{align}
 where the subscripts `$F$' and `$L$' refer to the properties calculated on the flight-side and lab-side, respectively. Equation \eqref{eq:INFER_FINAL_HF_HL_TRANSFORMATION} expresses $\hat{g}$ implicitly as the composition of the inverse map $H_F^{-1}$ with $e^{K_0} H_L$. 
Since $\chi(s) > 0$ on the interval $[g^{(w)}_F,1]$, the integral $H_F(s)$ is strictly monotonically decreasing and therefore the inverse $H_F^{-1}$ exists, and the formula \eqref{eq:INFER_FINAL_HF_HL_TRANSFORMATION} is well-posed. The monotonicity of the integrals $H_L, H_F$ for three different tests with varying wall and edge temperatures is shown in figure \ref{fig:all_H_pots}. \vspace{0mm}
{\color{DarkGreen}
We construct the inverse $H_F^{-1}$ numerically by tabulating $H_F$ on a monotone $g$-grid and interpolating $\hat{g}$ as a function of $H_F$. The numerical implementation of the $\hat{g}$-map proceeds as follows. Given the laboratory solution, the running integral $H_L(g) = \int_{g_\text{interior}}^1 \chi_{_{0}}(s){d}s$ is evaluated on the native $g$-grid of the lab profile. The lab-side property data and solution are readily available from \textsc{Plasflowsolver} \citep{lanza2025plasflowsolver}. Independently, the flight-side integral $H_F(\hat{g}) = 
\int_{\hat{g}_\text{interior}}^1 \chi_{_{\varepsilon}}({s}){d}{s}$ is tabulated on an arbitrary monotone ${g}$-grid using only the flight thermochemical property data from \textsc{Mutation++} \citep{mutation++ref}. It is important to stress that no explicit information about the shape of the flight temperature profile as a function of $\eta$ in the boundary layer is introduced in this step; only the functional dependence of the flight-side property $\chi_F$ on $g$ enters through $H_F$, where the role of the arbitrary g-grid is to define the discrete sampling on which the integral $H_F$ is tabulated. In other words, no knowledge of the flight temperature profile $\hat{g}(\eta)$ is required at this stage, as that is the solution we are seeking. For each interior lab point $g$, the target value $\mathcal{H}_L(g) = e^{K_0}H_L(g)$ is computed, and $\hat{g}$ is recovered by solving $H_F(\hat{g}) = \mathcal{H}_L(g)$, i.e.\ $\hat{g} = H_F^{-1}(\mathcal{H}_L(g))$. Since $\chi > 0$ everywhere, $H_F \text{ and } H_L$ are strictly monotonically decreasing, so this inversion is unique and is performed (easily) numerically by interpolating the pre-tabulated $\bigl(H_F,\,\hat{g}\bigr)$ table as: for any target value $\mathcal{H}_L(g)$ lying between two consecutive $H_F(\hat{g}_i)$ and $H_F(\hat{g}_{i+1})$, the correct mapped $\hat{g}$ must lie between $\hat{g}_i$ and $\hat{g}_{i+1}$, which is then obtained by interpolating between these neighboring tabulated points. See figure \ref{fig:illustrate_interp} for the illustration of the interpolation procedure. The construction guarantees that the wall and edge boundary conditions are recovered exactly: $g = g_w^L$ maps to $\hat{g} = g_w^F$ and $g = 1$ maps to $\hat{g} = 1$. The interior mapping is determined entirely by the thermochemical property variation ($\chi_L$ and $\chi_F$) through the integral invariant. For reproducibility and general use, the accompanying utility used to generate the $\hat{g}$-map and extrapolation results is publicly available at
\url{https://github.com/shbader/HyFLEX}.

\par
\begin{figure}
\hspace{-1.75cm}
    \centering
    \includegraphics[width=0.6\linewidth]{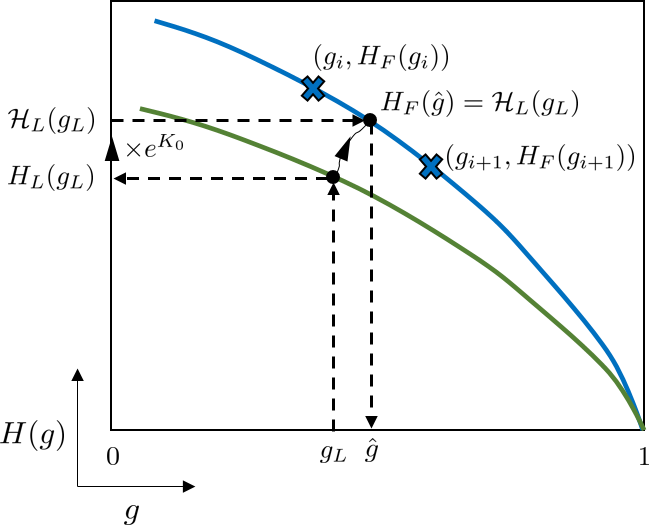}
    \caption{Illustration of the interpolation procedure used to infer $\hat{g}$ from the map \eqref{eq:INFER_FINAL_HF_HL_TRANSFORMATION}. Here the green curve represents the lab-side property integral $H_L$ and the blue curve represents its flight-side counterpart evaluated on an arbitrary grid. The mapping procedure follows a continuous path from the laboratory state to the flight state: starting at $g_L$, we compute the lab integral $H_L$, scale it to the flight target $\mathcal{H}_L = e^{K_0}H_L$, and find the corresponding flight temperature $\hat{g}$ through the inversion of the $H_F$ curve.}
    \label{fig:illustrate_interp}
\end{figure}
} 

Since the wall conditions are known for both the laboratory and flight cases, 
the constant coefficient $K_0$ in \eqref{eq:INFER_FINAL_HF_HL_TRANSFORMATION} 
is determined by evaluating the transformation at the wall:
\begin{align}\label{eq:K_0_formula}
    K_0 = \ln\!\left( \frac{H_F^{(w)}}{H_L^{(w)}} \right), \text{ with }    H_F^{(w)} = \int_{g_F^{(w)}}^{1} \chi^{}_{F}(s) ds, \text{ and }
    H_L^{(w)} = \int_{g_L^{(w)}}^{1} \chi^{}_{L}(s) ds.
\end{align}
In contrast, the group parameter $C_0$ corresponds to the $\eta$-stretch admitted by the equivalence group. For the tests discussed in the next section, we set $C_0=0$ to obtain the sufficient collapse, indicating the effective $\eta$ scale is already matched by how $\eta$ is defined. Also, from edge matching/normalization, we have $\hat{g}=1 \text{ when } g=1$; $\hat R_\rho(\hat g=1)= e^{2(A_0-C_0)\varepsilon}R_\rho = 1 \text{ when } R_{\rho}(g=1) = 1, \text{ and } \hat\ell_0(\hat g=1)=e^{(A_0+C_0)\varepsilon}\ell_0=1 \text{ when }\ell_{0}(g=1)=1$, this forces the scaling parameters $A_0 = C_0=0$. 
\par 
The non-trivial validation is that the resulting mapped profile $\hat g(\hat{\eta})$, together with the $\eta$-stretch, $\hat\eta=e^{C_0}\eta$, collapses with the independently computed flight solution $g_F(\eta)$, as discussed below in detail. The $g$-map obtained above is tested in three different stagnation point boundary layer flows with different wall $(T_w)$ and edge ($T_e$) temperatures for laboratory and flight cases. The idea is to test the robustness of the method in its ability to map the laboratory solution computed using modest values of $T_w \text{ and } T_e$ to a regime where the thermochemical property laws $(\text{e.g. } \chi)$ exhibit altered functional $g$-dependence due to dissociation and related effects. As stated above, we use \textsc{Plasflowsolver} \citep{lanza2025plasflowsolver} to compute the solutions of the original stagnation-point ODEs \eqref{eq:RElanzaODEs}. The gas mixture used for all numerical tests is \texttt{air$\_$11}, which comprises of  $\text{N}_2, \text{O}_2, \text{N}, \text{O}, \text{NO}, \text{N}^+, \text{O}^+, \text{NO}^+, \text{N}_2^+, \text{O}_2^+, \text{ and } \text{e}^-$ \citep{lanza2025plasflowsolver}.

\begin{figure}[h]
     \centering
     \begin{subfigure}[c]{0.8\textwidth}
         \centering
         \includegraphics[width=\textwidth]{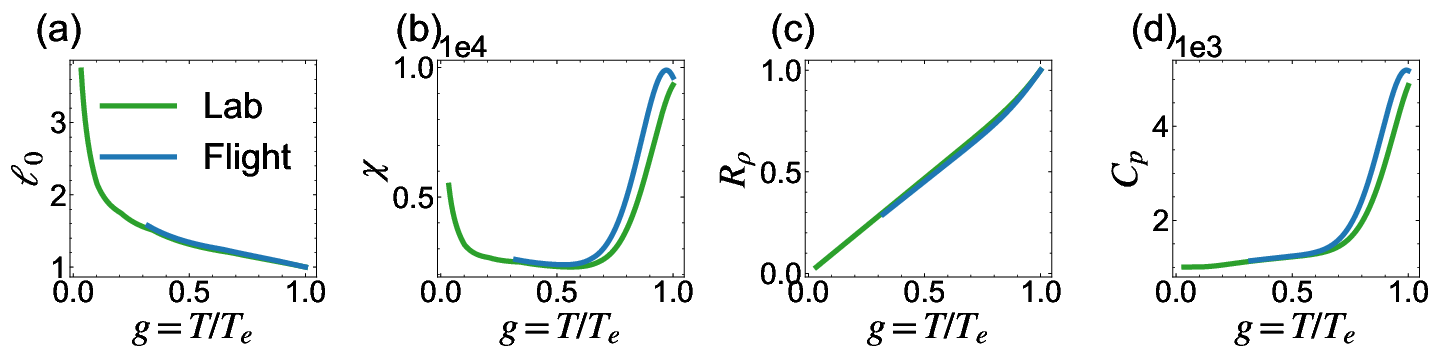}
         \label{fig:1g-mapping}
     \end{subfigure}
     \hfill 
     \begin{subfigure}[c]{0.18\textwidth}
         \centering
         \includegraphics[width=01\textwidth]{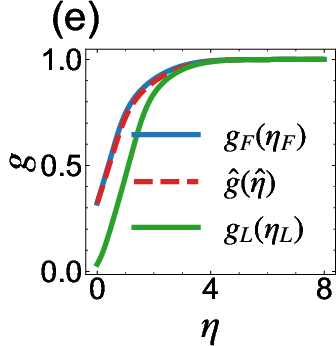}
         \label{fig:1properties-comparison}
     \end{subfigure} \vspace{-8.5mm} \\
     \begin{subfigure}[c]{0.8\textwidth}
         \centering
         \includegraphics[width=\textwidth]{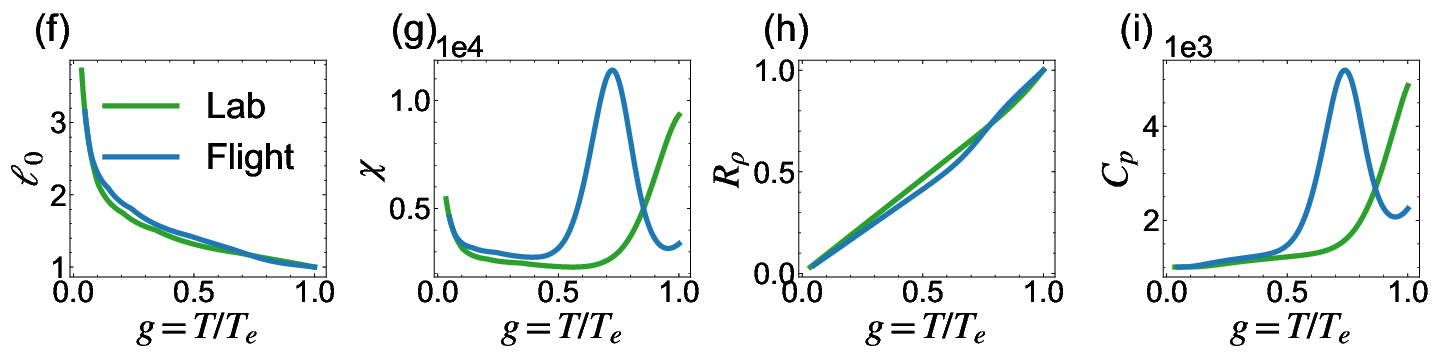}
         \label{fig:2g-mapping}
     \end{subfigure}
     \hfill 
     \begin{subfigure}[c]{0.18\textwidth}
         \centering
         \includegraphics[width=1\textwidth]{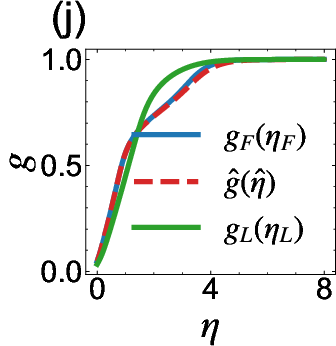}
         \label{fig:2properties-comparison}
     \end{subfigure} \vspace{-8.8mm} \\ 
     \begin{subfigure}[c]{0.8\textwidth}
         \centering
         \includegraphics[width=\textwidth]{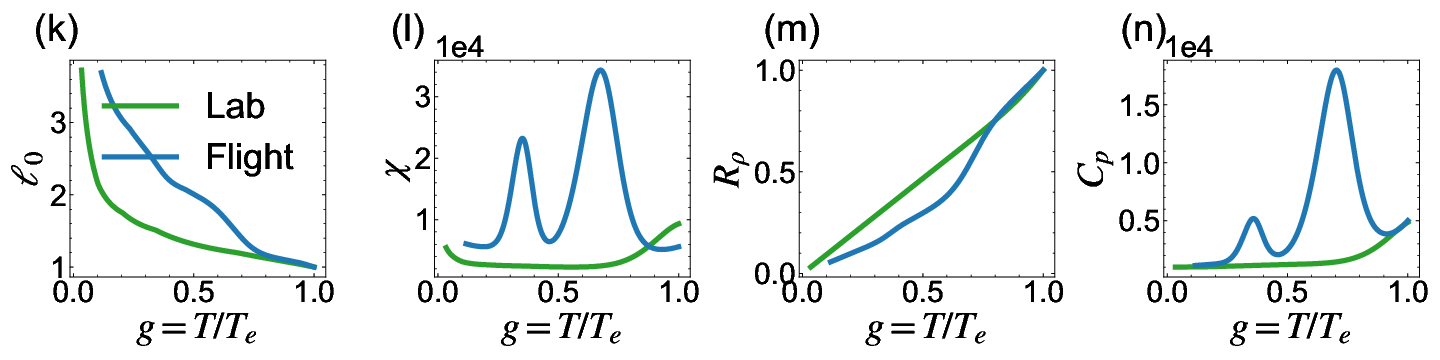}
         \label{fig:3g-mapping}
     \end{subfigure}
     \hfill 
     \begin{subfigure}[c]{0.18\textwidth}
         \centering
         \includegraphics[width=1\textwidth]{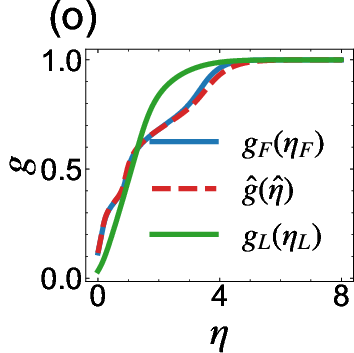}
         \label{fig:3properties-comparison}
     \end{subfigure}  \vspace{-8mm}
     \caption{
Thermophysical property profiles in laboratory and flight conditions and the corresponding boundary-layer temperature profiles for three test cases. Panels (a--d), (f--i), and (k--n) show the laboratory and flight laws for $\ell_0$, $\chi$, $R_{\rho}$, and $C_p$, respectively; panels (e), (j), and (o) compare the mapped temperature profile $\hat{g}(\hat{\eta})$ with the laboratory $g_L(\eta_L)$ and flight $g_F(\eta_F)$ profiles.
Row~1: $T_w^L=100\,\mathrm{K}$, $T_w^F=1000\,\mathrm{K}$, $T_e^L=2935\,\mathrm{K}$, $T_e^F=3103\,\mathrm{K}$.
Row~2: $T_w^L=100\,\mathrm{K}$, $T_w^F=200\,\mathrm{K}$, $T_e^L=2935\,\mathrm{K}$, $T_e^F=4165\,\mathrm{K}$.
Row~3: $T_w^L=100\,\mathrm{K}$, $T_w^F=1000\,\mathrm{K}$, $T_e^L=2935\,\mathrm{K}$, $T_e^F=8586\,\mathrm{K}$.
}
     \label{fig:all-overall-comparison}
\end{figure}

In figures \ref{fig:all-overall-comparison} (a-e), the wall temperature in the laboratory test is held fixed at $T_L^w=100 \text{ K }$, and that for the reference flight case, $T_F^w=1000 \text{ K }$. In the solver \citep{lanza2025plasflowsolver}, the target wall-heat flux $q_w= \sqrt{\tfrac{2\beta}{\mu_e\rho_e}}T_e\rho_w\lambda_w \left ( \tfrac{dg}{d\eta} \right )_{\eta=0}$ can be prescribed and the edge temperature $T_e$ is iterated until the specified target wall-heat flux is reached. In the cases shown in figures \ref{fig:all-overall-comparison}(a-e), the same value of $q_w$ is imposed for the laboratory and the reference flight cases, with laboratory case edge temperature $T_L^e = 2935 \text{ K }$ and the iterated flight side edge temperature $T_F^e = 3103 \text{ K }$. The stagnation pressure $(p_s)$ and the velocity gradient $\beta=(du_e/dx)_{x=0} = u/R^{*}_m $, where $u$ is the characteristic mixture freestream velocity and $R_m^{*}$ is the external radius of the probe/body, are matched while as the thermochemical edge states characterized by total enthalpy $(h_0)$ differ by $10-15\%$ between the laboratory and flight cases. 
To completely determine the stagnation-point heat flux in a ground test, LHTS \citep{kolesnikov1_1993conditions, kolesnikov2_2000concept} requires matching the parameters $(h_0, \beta, p_s)$ at the edge of the boundary layer together with the thermal boundary condition at the wall $(T_w)$. {\color{blue}In the present case (see figure \ref{fig:all-overall-comparison})(a-e), although the laboratory and flight configurations employ distinct wall thermochemical states ($T_w^L = 100~\mathrm{K}$, $T_w^F = 1000~\mathrm{K}$) and $h_0$ differs by $\sim 10-15\%$, enforcement of identical wall heat flux $q_w$ with different wall and edge states demonstrates that the LHTS matching conditions are only \textit{sufficient} (but not \textit{necessary}) conditions of equality of wall-heat fluxes between the laboratory test and flight. In the equivalence framework, we observe that even when $T_w$ differs significantly and $h_0$ mismatches, there exists a nontrivial transformation in the equivalence class of the underlying ODEs that maps $g$ from laboratory to flight and preserves the near-wall heat transfer. The classical LHTS framework is recovered as the mathematically trivial (identity) case of this mapping, corresponding to exact matching of the wall boundary condition $T_w$ and the edge thermochemical state $h_0$, and leading to identical stagnation-point boundary-layer temperature profiles in the laboratory and flight configurations. It must be stressed here that while LHTS is a degenerate case from a mathematical standpoint, its experimental utility is nontrivial. In this sense, the present analysis provides a group-theoretic interpretation of LHTS by identifying its matching conditions as the identity limit of a broader equivalence mapping between laboratory and flight stagnation-point similarity solutions.}


In Figures \ref{fig:all-overall-comparison}(f-j) and \ref{fig:all-overall-comparison}(k-o), two additional tests are presented with different wall temperatures $(T_L^w, T_F^w)$ and varying $q_w$ between the laboratory and reference flight cases. Despite these variations, the mapped temperature profile $\hat{g}$ exhibits a good collapse with the flight solution $g_F$ across the boundary layer. {\color{blue}It must be noted that the collapse is not exact and any departures from the exact collapse of mapped profile $\hat g$ with the flight profile $g_F$ may be attributed to the invalidity of the integral invariant condition in the outer part of the boundary layer ($\eta > 3$). The constant $K_0$ in \eqref{eq:K_0_formula} is obtained by using the wall condition and it is used for scaling the $H_L(g)$ integral throughout the layer. For example, in Figures \ref{fig:all_H_pots}(b,c), there are regions in the boundary layer when the validity of this invariant breaks; these regions are marked by $H_F(g) < H_L(g)$ which contradicts the positive scaling factor $K_0$ set by the wall condition. This is why the profiles are more accurate in the inner region of the boundary layer, $\eta \leq 3$ in Figures \ref{fig:all-overall-comparison}(j,o). The structure of the mapped profile is accurate close to the wall, giving a correct gradient and consequently an accurate measure of heat-transfer. Overall, the derived integral invariant yields an exact temperature profile within the interpolation error in the inner part of the boundary layer, $\eta \leq 3$, and is able to successfully incorporate the effects of thermochemical property variation on the temperature profile.} 
\par 

{\color{magenta}
The construction and the associated results above determine the nonlinear temperature reparametrization $g\mapsto\hat g$ from the $\chi$-law. Once $\hat g$-map is inferred, the other arbitrary elements are constrained. This prevents the physical flight property data to be an exact transformed image of the laboratory data for every arbitrary element. As discussed above, under the edge normalization, we have $\hat{g}=1 \text{ when } g=1$; $\hat R_\rho(\hat g=1)= 1 \text{ when } R_{\rho}(g=1) = 1, \text{ and } \hat\ell_0(\hat g=1)=1 \text{ when }\ell_{0}(g=1)=1$, this forces the scaling parameters $A_0 = C_0=0$, reducing the $\ell_0$ and $R_\rho$ laws to $\hat R_\rho(\hat g)=R_{\rho}(g), \hat\ell_0(\hat g)=\ell_{0}(g)$. The nontrivial freedom in the thermal map is carried by the Jacobian-weighted transformations of $\chi$ and $C_p$. Since the $\chi$ law satisfies the $\hat{g}$-map by construction, the $C_p$ law is then compatible with the same map if $\hat{C}_{p}(\hat g)\simeq e^{K_0}J^{-1}C_{p,L}(g) = ({\chi_F(\hat g)}/{\chi_L(g)})C_{p,L}(g)$. As an example, for the cases presented in rows 2 and 3 of the Figure \ref{fig:all-overall-comparison}, $C_p$ remains comparatively compatible with the $\chi$-inferred map. The results are shown in the figure \ref{fig:case2_case3_CP_compare}. The incompatibility of the $R_\rho$ and $\ell_0$ transformations do not theoretically invalidate the equivalence transformation: the Lie map remains exact for the transformed member of the differential-equation class. Rather, they quantify the extent to which the prescribed physical flight property data departs from the exact transformed counterpart.

\begin{figure}
    \centering
    \includegraphics[width=0.8\linewidth]{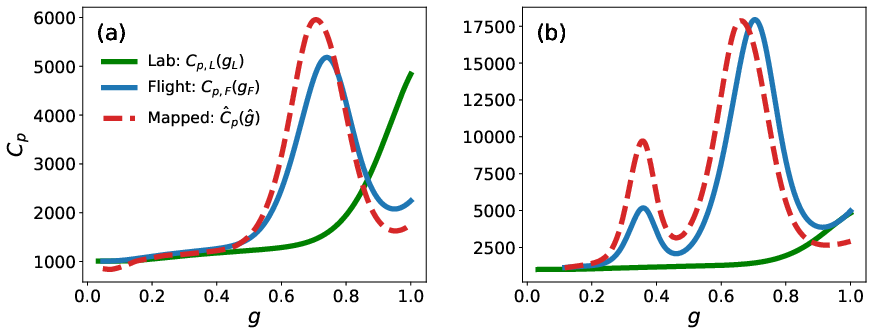}
    \caption{Comparison of transformed $\hat{C}_p$ obtained from the $\chi$-based reparametrization of temperature for the two cases presented in the rows 2 (panel (a) here) and 3 (panel (b) here) of figure \ref{fig:all-overall-comparison}.}
    \label{fig:case2_case3_CP_compare}
\end{figure}
}

\section{Conclusions}
To address the problem of lab-to-flight extrapolation in high-speed, dissociative boundary layers, we present the Lie equivalence symmetry analysis of hypersonic stagnation-point ODEs to derive mappings $(\hat g)$ for non-dimensional temperature $(g)$. The framework extends beyond classical similitude by introducing invariants associated with the underlying equivalence group, which can, in principle, include the similarity parameters employed in classical similitude as a subset. The practical consequence of this extension alleviates the need to match the entire set of non-dimensional groups characteristic of a set of governing equations. Instead, by allowing the property laws to transform, the invariants allow for non-linear maps which can be used to infer the laboratory-to-flight mappings. In practice, implementing these maps requires only the laboratory-scale solution together with lab- and flight-side thermochemical property data; for commonly used thermochemical databases, these properties are available up to temperatures of order $2\times 10^{4}\,\mathrm{K}$ \citep{mutation++ref}. For all the cases considered in this study, the derived $g$-map demonstrates a good collapse with the \textcolor{blue}{corresponding similarity-ODE flight solution}, verifying the validity of the derived invariant. 


\textcolor{blue}{A limitation of the present study is that the analysis is carried out for a system of coupled, similarity-reduced stagnation-point boundary-layer ODEs. The results should therefore be interpreted as a proof of concept for this reduced formulation, rather than as a validation of a general laboratory-to-flight extrapolation framework. Accordingly, the invariant derived here is valid for the specific ODE class considered, and its numerical assessment is performed against independently computed similarity-ODE flight solutions. We have not considered the full boundary-layer PDE system for a more general equivalence mapping, or validated our current results against multidimensional CFD, or experimental measurements. Despite these limitations, the similarity-reduced stagnation point ODEs still provide a practically accurate representation of the stagnation region in hypersonic flows.}

Overall, the Lie equivalence-symmetry framework provides a systematic means of generalizing local similarity concepts for hypersonic boundary layers. Extension of the present analysis to incorporate additional physical complexity, including chemical and thermal nonequilibrium and radiative effects, will be pursued in future editions of this work.


\begin{bmhead}[Acknowledgment]
The authors gratefully acknowledge Domenico Lanza (Politecnico di Milano),  Massimo Franco (UIUC) and Francesco Panerai (UIUC)  for fruitful discussions and their assistance with \textsc{PlasFlowSolver}.
\end{bmhead}

\begin{bmhead}[Funding]
S.B. gratefully acknowledges financial support from CHESS, the Center for Hypersonics and Entry System Studies, in the Grainger College of Engineering at the University of Illinois Urbana-Champaign.
\end{bmhead}

\begin{bmhead}[Declaration of interests]
The authors report no conflict of interest.
\end{bmhead}

\begin{bmhead}[Author contributions]
SB:  Conceptualization, Methodology, Data curation, Software, Formal analysis, Validation, Writing – original draft, review \& editing. DB: Supervision, Resources,  Project administration, Funding acquisition, Writing – review \& editing.
\end{bmhead}


\begin{appen}
{
\color{DarkGreen}

\section{Lie symmetry and equivalence transformations}\label{APPEND-A-LIETHEORY}
A Lie \emph{point} symmetry of a differential equation is a smooth change of variables that maps every solution of the equation to another solution. Following \citet{hydon2000} and \citet{cantwell2002introduction}, one regards a symmetry as a diffeomorphism
\(\Gamma : (x,y) \mapsto (\hat{x},\hat{y})\) which induces a corresponding action on all derivatives \(y^{(k)}\). When the transformed derivatives \(\hat{y}^{(k)}\) are substituted into the differential equation, the equation retains its form; in other words, the solution set is invariant under \(\Gamma\). These continuous symmetries form a Lie group, and their infinitesimal generators provide a systematic way to construct transformations that can simplify, reduce, or relate solutions of differential equations. For brevity, we will mainly follow the simple notation of \citet{hydon2000} to discuss the Lie point symmetries in the context of an ODE of any arbitrary order `$n$'. Consider an $n$th-order ODE of the form,
\begin{align}\label{eq:LS_ODE_nth_lie_symm}
 y^{(n)} = \mathcal{F}\!\left(x, y, y^{(1)}, \ldots y^{(n-1)}\right), \quad
y^{(k)} \equiv \frac{d^{k}y}{dx^{k}} \quad k = 1, 2, \ldots n.
\end{align}
For $\varepsilon$ sufficiently close to zero, the prolonged Lie symmetries are of the form,
\begin{subequations}
    \begin{align}
\hat x &= x + \varepsilon \xi + \mathcal{O}(\varepsilon^{2})\\
\hat y &= y + \varepsilon \eta + \mathcal{O}(\varepsilon^{2})\\
\hat y^{(k)} &= y^{(k)} + \varepsilon \eta^{(k)} + \mathcal{O}(\varepsilon^{2}),
\quad k \ge 1
\end{align}
\end{subequations}
where
\begin{align}
\left(\xi(x, y), \eta(x, y)\right)
= \left(
\left.\frac{d\hat x}{d\varepsilon}\right|_{\varepsilon=0},
\left.\frac{d\hat y}{d\varepsilon}\right|_{\varepsilon=0}
\right), 
\end{align}
is the tangent vector at $(x,y)$, and $\eta^{(k)}$ can be obtained recursively as, 
\begin{align}
    \eta^{(k)}\bigl(x, y, y', \ldots y^{(k)}\bigr)
= D_x \eta^{(k-1)} - y^{(k)} D_x \xi, 
\end{align}
$\text{ with } \eta^{(1)} = D_x \eta - y^{(1)} D_x \xi, \text{ and } D_x = \partial_x + y^{(1)} \partial_y + y^{(2)} \partial_{y^{(1)}} + \cdots$. The superscripts $(1), (2), \ldots (k)$ on variable $y$ represent the derivatives. For $\eta$ as in $\eta^{(1)}$, it is merely an index and it does not represent the derivative of $\eta$. 
The action of $\Gamma$ induces an action on the derivatives $y^{(k)}$, yielding the map, $\Gamma : (x, y, y', \ldots y^{(n)}) \mapsto (\hat{x}, \hat{y}, \hat{y}', \ldots \hat{y}^{(n)})$ with $\hat y^{(k)} = {d^{k}\hat y}/{d\hat x^{k}}, k = 1 \ldots n$.
To describe how Lie symmetries act on derivatives up to order $n$ we define the \emph{prolonged infinitesimal generator}
\begin{align}
X^{(n)} &= \xi \partial_x + \eta \partial_y + \eta^{(1)} \partial_{y^{(1)}} + \cdots + \eta^{(n)} \partial_{y^{(n)}}.
\end{align}
Here the coefficient of $\partial_{y^{(k)}}$ is the $\mathcal{O}(\varepsilon)$ term in the expansion of $\hat y^{(k)}$ so $X^{(n)}$ represents the tangent vector in the space of variables $(x, y, y^{(1)}, \ldots y^{(n)})$. Using this prolonged generator the linearized symmetry condition can be written compactly as
\begin{align}\label{eq:LS_ODE_LSC}
X^{(n)} [\Delta] &= 0,
\quad \text{when \eqref{eq:LS_ODE_nth_lie_symm} holds}.
\end{align}
where
\begin{align}\label{eq:STANDARD_FORM_DELTA_DEFINITION}
    \Delta := y^{(n)} - \mathcal{F}\!\left(x, y, y^{(1)}, \ldots y^{(n-1)}\right)=0.
\end{align}
The linearized symmetry condition \eqref{eq:LS_ODE_LSC} is split into a system of determining equations whose solution leads to the Lie group of symmetries of the given ODE \eqref{eq:LS_ODE_nth_lie_symm}. For more details, the interested reader is referred to \citet{hydon2000} and \citet{cantwell2002introduction}.  

Lie point symmetries act on the variables of a single, fixed differential equation while keeping all its coefficients (like viscosity, conductivity etc.) and constitutive laws fixed. However, these arbitrary functions like coefficients and constitutive laws may also be allowed to transform. This results in symmetries acting on the entire class of equations. Simply put, if we include the arbitrary functions in the process of obtaining symmetries as outlined above, the resulting symmetries are known as Lie \textit{equivalence} symmetries. The general framework for obtaining equivalence symmetries can be outlined as,
\begin{align}\label{eq:EQUIV-LS_ODE_nth_lie_symm}
\mathcal{F}\!\left(x, y, y^{(1)}, \ldots, y^{(n-1)}, y^{(n)}, K(y), K^{(1)}(y), \ldots, K^{(n-1)}(y), K^{(n)}(y) \right)=0,
\end{align}
The dependence of the arbitrary function $K(y)$ on the independent variable is given by the auxiliary equation, ${\partial K}/{\partial x} = 0$. For the case of multiple variables, it could be a system of equations. The simple case of the arbitrary function depending only on the variable $y$ is demonstrated in this example. Multiple variables may exist with multiple dependencies, as will be the case in stagnation-point ODEs. In the enlarged system with arbitrary function(s) as extra dependent variable(s), the prolonged infinitesimal generator includes additional components,
\begin{align}
X^{(n)} &= \xi \partial_x + \eta \partial_y + \eta^{(1)} \partial_{y^{(1)}} + \cdots + \eta^{(n)} \partial_{y^{(n)}} + \kappa \partial_K + \kappa^{(1)} \partial_{K^{(1)}} + \ldots + \kappa^{(n)} \partial_{K^{(n)}}.
\end{align}
The rest of the procedure is similar: the prolonged infinitesimal generator is applied to the auxilliary system of equations in addition to the main ODE, and the resulting system of determining equations is solved for the equivalence group. For more details, the interested reader is referred to \citet{ovsiannikov2014group}.

\section{Invariance check of the derived transformations}\label{APPEND_INV_CHECK}

As a sanity check, we verify that the transformations obtained
from the equivalence generators preserve the form
of each equation in the stagnation-point ODE system. The governing equations are,
\begin{align}
    \frac{dV}{d\eta}+F &= 0,
    \label{eq:app_continuity}
    \\
    V\frac{dF}{d\eta}
    &=
    \frac{1}{2}\left(R_\rho-F^2\right)
    +
    \frac{d}{d\eta}
    \left(
        \ell_0\frac{dF}{d\eta}
    \right),
    \label{eq:app_momentum}
    \\
    V\frac{dg}{d\eta}
    &=
    \frac{1}{C_p}
    \frac{d}{d\eta}
    \left(
        \chi\frac{dg}{d\eta}
    \right).
    \label{eq:app_energy}
\end{align}

The infinitesimal generator is,
\begin{align}
X
&=
C_0\eta\frac{\partial}{\partial \eta}
+
(A_0-C_0)F\frac{\partial}{\partial F}
+
A_0V\frac{\partial}{\partial V}
+
\phi_g(g)\frac{\partial}{\partial g}
+
2(A_0-C_0)R_\rho\frac{\partial}{\partial R_\rho}
+
(A_0+C_0)\ell_0\frac{\partial}{\partial \ell_0}
\notag\\
&\quad
+
\left[
K_0-\frac{d}{dg}\left(\phi_g(g)\right)
\right]\chi\frac{\partial}{\partial \chi}
+
\left[
K_0-A_0-C_0-\frac{d}{dg}\left(\phi_g(g)\right)
\right]C_p\frac{\partial}{\partial C_p}.
\end{align}

The corresponding finite transformations are,
\begin{align}\label{app:nFV_transf}
\hat\eta = e^{C_0\varepsilon}\eta, \quad \hat V = e^{A_0\varepsilon}V, \quad \hat F = e^{(A_0-C_0)\varepsilon}F,
\end{align}
together with,
\begin{align}\label{app:jacobian}
\frac{\partial \hat g}{\partial \varepsilon} = \phi_g(\hat g), \quad \hat g(0,g)&=g, \text{ and } J = \frac{d\hat{g}}{dg}.
\end{align}\label{app:arbitrary_transf}
The arbitrary elements transform as,
\begin{align}
\hat R_\rho(\hat g)
&=
e^{2(A_0-C_0)\varepsilon}R_\rho(g), \label{appA_Rlaw} \\
\hat\ell_0(\hat g)
&=
e^{(A_0+C_0)\varepsilon}\ell_0(g), \label{appA_l0law} \\
\hat\chi(\hat g)
&=
e^{K_0\varepsilon}
J^{-1}
\chi(g), \label{appA_chilaw} \\
\hat C_p(\hat g)
&=
e^{(K_0-A_0-C_0)\varepsilon}
J^{-1}
C_p(g).\label{appA_cplaw}
\end{align}

Since \(\hat\eta=e^{C_0\varepsilon}\eta\), the derivative operator transforms as,
\begin{equation}
    \frac{d}{d\hat\eta}
    =
    e^{-C_0\varepsilon}\frac{d}{d\eta}.
    \label{eq:app_derivative_transform}
\end{equation}

\subsection{Continuity equation}
\begin{align}
    \frac{d\hat V}{d\hat\eta}
    &=
    e^{-C_0\varepsilon}
    \frac{d}{d\eta}
    \left(
        e^{A_0\varepsilon}V
    \right)
    \notag \\
    &=
    e^{-C_0\varepsilon}e^{A_0\varepsilon}
    \frac{dV}{d\eta}
    \notag \\
    &=
    e^{(A_0-C_0)\varepsilon}
    \frac{dV}{d\eta}.
    \label{eq:app_cont_derivative}
\end{align}
Since $\hat F=e^{(A_0-C_0)\varepsilon}F$, the transformed continuity equation is,
\begin{align}
    \frac{d\hat V}{d\hat\eta}+\hat F
    &=
    e^{(A_0-C_0)\varepsilon}
    \left(
        \frac{dV}{d\eta}+F
    \right)
    =0. \notag \\
    \Rightarrow \frac{d\hat V}{d\hat\eta}+\hat F
    &=0.
    \label{eq:app_cont_invariance}
\end{align}
Thus, if \eqref{eq:app_continuity} is satisfied, then the transformed
continuity equation is also satisfied.

\subsection{Momentum equation}

First,
\begin{align}
    \frac{d\hat F}{d\hat\eta}
    &=
    e^{-C_0\varepsilon}
    \frac{d}{d\eta}
    \left(
        e^{(A_0-C_0)\varepsilon}F
    \right)
    \notag \\
    &=
    e^{(A_0-2C_0)\varepsilon}
    \frac{dF}{d\eta}.
    \label{eq:app_F_derivative}
\end{align}
Therefore the transformed left-hand side is,
\begin{align}
    \hat V\frac{d\hat F}{d\hat\eta}
    &=
    \left(e^{A_0\varepsilon}V\right)
    \left(
        e^{(A_0-2C_0)\varepsilon}\frac{dF}{d\eta}
    \right)
    \notag \\
    &=
    e^{2(A_0-C_0)\varepsilon}
    V\frac{dF}{d\eta}.
    \label{eq:app_mom_lhs}
\end{align}
The first term on the right-hand side transforms as,
\begin{align}
    \frac{1}{2}
    \left(
        \hat R_\rho-\hat F^2
    \right)
    &=
    \frac{1}{2}
    \left(
        e^{2(A_0-C_0)\varepsilon}R_\rho
        -
        e^{2(A_0-C_0)\varepsilon}F^2
    \right)
    \notag \\
    &=
    e^{2(A_0-C_0)\varepsilon}
    \frac{1}{2}
    \left(
        R_\rho-F^2
    \right).
    \label{eq:app_mom_pressure}
\end{align}
For the second term on the right-hand side, we have,
\begin{align}
    \hat\ell_0\frac{d\hat F}{d\hat\eta}
    &=
    \left(
        e^{(A_0+C_0)\varepsilon}\ell_0
    \right)
    \left(
        e^{(A_0-2C_0)\varepsilon}\frac{dF}{d\eta}
    \right)
    \notag \\
    &=
    e^{(2A_0-C_0)\varepsilon}
    \ell_0\frac{dF}{d\eta}.
    \label{eq:app_mom_flux}
\end{align}
Hence,
\begin{align}
    \frac{d}{d\hat\eta}
    \left(
        \hat\ell_0\frac{d\hat F}{d\hat\eta}
    \right)
    &=
    e^{-C_0\varepsilon}
    \frac{d}{d\eta}
    \left(
        e^{(2A_0-C_0)\varepsilon}
        \ell_0\frac{dF}{d\eta}
    \right)
    \notag \\
    &=
    e^{2(A_0-C_0)\varepsilon}
    \frac{d}{d\eta}
    \left(
        \ell_0\frac{dF}{d\eta}
    \right).
    \label{eq:app_mom_diffusion}
\end{align}
Combining \eqref{eq:app_mom_lhs}--\eqref{eq:app_mom_diffusion}, we can write,
\begin{align}
    \hat{V}\frac{d\hat{F}}{d\hat{\eta}}
    -
    \frac{1}{2}
    \left(
        \hat{R}_\rho-\hat{F}^2
    \right)
    -
    \frac{d}{d\hat{\eta}}
    \left(
        \hat{\ell}_0\frac{d\hat{F}}{d\hat{\eta}}
    \right)
    &=
    e^{2(A_0-C_0)\varepsilon}
    \left[
    {V}\frac{d{F}}{d{\eta}}
    -
    \frac{1}{2}
    \left(
        {R}_\rho-{F}^2
    \right)
    -
    \frac{d}{d{\eta}}
    \left(
        {\ell}_0\frac{d{F}}{d{\eta}}
    \right)
    \right]
    =0. \notag \\
    \Rightarrow \hat{V}\frac{d\hat{F}}{d\hat{\eta}}
    &=
    \frac{1}{2}
    \left(
        \hat{R}_\rho-\hat{F}^2
    \right)
    +
    \frac{d}{d\hat{\eta}}
    \left(
        \hat{\ell}_0\frac{d\hat{F}}{d\hat{\eta}}
    \right).
    \label{eq:app_mom_invariance}
\end{align}
Thus, when \eqref{eq:app_momentum} is satisfied, the transformed momentum equation is also satisfied.

\subsection{Energy equation}
We have,
\begin{align}
    \frac{d\hat g}{d\eta}
    =
    J\frac{dg}{d\eta},
    \qquad
    J=\frac{\partial \hat g}{\partial g}.
    \label{eq:app_g_derivative_eta}
\end{align}
Therefore,
\begin{align}
    \frac{d\hat g}{d\hat\eta}
    =
    e^{-C_0\varepsilon}
    J\frac{dg}{d\eta}.
    \label{eq:app_g_derivative_hateta}
\end{align}
The transformed left-hand side of the energy equation is then,
\begin{align}
    \hat V\frac{d\hat g}{d\hat\eta}
    &=
    \left(
        e^{A_0\varepsilon}V
    \right)
    \left(
        e^{-C_0\varepsilon}
        J\frac{dg}{d\eta}
    \right)
    \notag \\
    &=
    e^{(A_0-C_0)\varepsilon}
    J V\frac{dg}{d\eta}.
    \label{eq:app_energy_lhs}
\end{align}

Using \eqref{appA_chilaw} and
\eqref{eq:app_g_derivative_hateta},
\begin{align}
    \hat\chi\frac{d\hat g}{d\hat\eta}
    &=
    \left(
        e^{K_0\varepsilon}J^{-1}\chi
    \right)
    \left(
        e^{-C_0\varepsilon}
        J\frac{dg}{d\eta}
    \right)
    \notag \\
    &=
    e^{(K_0-C_0)\varepsilon}
    \chi\frac{dg}{d\eta}.
    \label{eq:app_energy_flux}
\end{align}
Differentiating the above with respect to $\hat\eta$ gives,
\begin{align}
    \frac{d}{d\hat\eta}
    \left(
        \hat\chi\frac{d\hat g}{d\hat\eta}
    \right)
    &=
    e^{-C_0\varepsilon}
    \frac{d}{d\eta}
    \left[
        e^{(K_0-C_0)\varepsilon}
        \chi\frac{dg}{d\eta}
    \right]
    \notag \\
    &=
    e^{(K_0-2C_0)\varepsilon}
    \frac{d}{d\eta}
    \left(
        \chi\frac{dg}{d\eta}
    \right).
    \label{eq:app_energy_flux_derivative}
\end{align}
From \eqref{appA_cplaw},
\begin{align}
    \frac{1}{\hat C_p}
    =
    e^{-(K_0-A_0-C_0)\varepsilon}
    J\frac{1}{C_p}.
    \label{eq:app_inverse_Cp}
\end{align}
Thus the transformed right-hand side is,
\begin{align}
    \frac{1}{\hat C_p}
    \frac{d}{d\hat\eta}
    \left(
        \hat\chi\frac{d\hat g}{d\hat\eta}
    \right)
    &=
    \left(
        e^{-(K_0-A_0-C_0)\varepsilon}
        J\frac{1}{C_p}
    \right)
    \left(
        e^{(K_0-2C_0)\varepsilon}
        \frac{d}{d\eta}
        \left(
            \chi\frac{dg}{d\eta}
        \right)
    \right)
    \notag \\
    &=
    e^{(A_0-C_0)\varepsilon}
    \frac{J}{C_p}
    \frac{d}{d\eta}
    \left(
        \chi\frac{dg}{d\eta}
    \right).
    \label{eq:app_energy_rhs}
\end{align}
Hence the transformed energy equation becomes,
\begin{align}
  \hat{V}\frac{d\hat{g}}{d\hat{\eta}}
    -
    \frac{1}{\hat{C}_p}
    \frac{d}{d\hat{\eta}}
    \left(
        \hat{\chi}\frac{d\hat{g}}{d\hat{\eta}}
    \right)
    &=
     e^{(A_0-C_0)\varepsilon}
     J
     \left[
     V\frac{dg}{d\eta}
    -
    \frac{1}{C_p}
    \frac{d}{d\eta}
    \left(
        \chi\frac{dg}{d\eta}
    \right)
    \right]
    =0, \notag \\
    \Rightarrow \hat{V}\frac{d\hat{g}}{d\hat{\eta}}
    &=
    \frac{1}{\hat{C}_p}
    \frac{d}{d\hat{\eta}}
    \left(
        \hat{\chi}\frac{d\hat{g}}{d\hat{\eta}}
    \right).
    \label{eq:app_energy_transformed}
\end{align}
Since $J\neq 0$ for an invertible finite temperature map $d\hat{g}=Jdg$, when \eqref{eq:app_energy} is satisfied, the transformed energy equation is also satisfied.
}

\end{appen}

\bibliographystyle{jfm}
\bibliography{jfm}



\end{document}